\documentclass[10pt]{article}

\begin{document}

\title{The 2 - particle irreducible effective action in gauge theories}
\author{E. A. Calzetta\thanks{%
Email: calzetta@df.uba.ar} \\
Departamento de Fisica, Facultad de Ciencias Exactas y Naturales\\
Universidad de Buenos Aires- Ciudad Universitaria,\\
1428 Buenos Aires, Argentina}
\date{Feb 18, 2004}
\maketitle

\begin{abstract}
The goal of this paper is to develop the formalism of the two-particle
irreducible (2PI) \cite{LW61} (or Cornwall - Jackiw - Tomboulis (CJT) \cite
{CJT}) effective action (EA) in a way appropiate to its application to non
equilibrium gauge theories. We hope this review article will stimulate new
work into this field.
\end{abstract}

\section{Introduction}

The goal of this paper is to develop the formalism of the two-particle
irreducible (2PI) \cite{LW61} (or Cornwall - Jackiw - Tomboulis (CJT) \cite
{CJT}) effective action (EA) in a way appropiate to its application to non
equilibrium gauge theories. The usual formulation of the 2PIEA cannot be
extended to these theories because of the special features of gauge
invariance. There are two difficulties in particular which require
consideration, namely, the existence of constraints linking the Schwinger
functions of the theory among themselves \cite{Emil}, and the peculiarities
of the gauge - fixing procedure. The former are expressed by the so-called
Takahashi - Ward or Slavnov - Taylor identities, which in turn derive from
the Zinn - Justin equation (see below) \cite{ZJ}. The latter raises the
issue, exclusive to gauge theories, of the gauge fixing dependence of
theoretical constructs \cite{gaugedependence}. A clear understanding of this
issue is essential to the physical interpretation of the theory.

Non equilibrium quantum field theory \cite{NEQFT} has evolved in the last
few years mostly in the context of non - gauge field theories. In these
studies, two tools have proved extremely valuable, namely the closed - time
- path (CTP), IN-IN or Schwinger - Keldish formalism \cite{inin}, and the
2PI or CJT EA. The former allows us to study the causal evolution of quantum
fields, in contradistinction to the IN -OUT or mixed advanced and retarded
boundary conditions appropiate to the study of scattering processes. The
latter provides a comprehensive framework where various non - perturbative
approaches may be most efficiently implemented. The need to go beyond simple
- minded perturbation theory is an universal feature of high energy
nonequilibrium processes in non linear theories.

When we survey the literature on gauge theories, we find that the one -
particle ireducible (1PI) IN - OUT EA is a well developed tool which has
found its way into most modern textbooks\cite{Weinberg2,Peskin}. The CTP
formulation of gauge theories, although less widespread, has also been the
subject of several important investigations and may be considered well
understood \cite{Geiger,Son}. The 2PIEA, on the other hand, has only
recently come under study \cite{Emil,Gabor,Jurgen}. We hope this review
article will stimulate new work into this field.

Of course, the subject of non equilibrium gauge theories is so vast that it
becomes impossible to make progress without some well defined self -
impossed limitations from the outset. We will restrict ourselves to Yang -
Mills and to non linear abelian theories such as QED and SQED. We shall make
no explicit attempt to discuss gravity, form fields or string theories \cite
{Weinberg3}.

These self - imposed limitations in aims are correlated with some necessary
a priori technical choices. We shall discuss only the path integral Fadeev -
Popov quantization of gauge theories. Although we shall use BRST invariance
at several stages, we shall not apply methods such as BRST or Batalin -
Vilkovisky quantization, which really come on their own only in more
demanding applications \cite{Weinberg2}. We shall use DeWitt's notation and
are deeply in debt to DeWitt's insights \cite{DeWitt}, but we shall not use
the gauge - independent formulation of DeWitt and Vilkovisky \cite{VDW}(on
this subject, see the discussion in \cite{KKR}), nor more recent
developments by DeWitt and collaborators \cite{Carmen}.

When gauge symmetries are unbroken, there are no preferred directions in
gauge space, and all background fields will vanish identically. Therefore,
the only degrees of freedom in the 2PI formalism shall be the propagators or
two - point functions. Also, there will be no need to distinguish between
the usual and the DeWitt-Abbott gauge invariant EA \cite{GIEA}, nor to
introduce gauge fixing conditions appropiate to the study of broken gauge
theories, such as the $R\xi $ family of gauges \cite{Weinberg2}. We shall
only assume that the gauge fixing condition is linear on the quantum fields.
On the other hand, we shall be completely general regarding group structure,
matter content, (linear) gauge fixing condition and gauge fixing parameter.

The main results of this paper are formulae (\ref{TPIEA}) and (\ref{gamma2})
providing the definition of the 2PIEA for gauge theories. We shall then use
this construction to discuss the Zinn-Justin identities and the gauge
dependence of the propagators. These are in some sense known results, and we
include them to help the reader to connect the 2PI to earlier formulations
of non equilibrium field theories, and to better appreciate its power.

The paper is organized as follows. Section 2 is a review of the path
integral approach to gauge theories, including the IN-OUT effective action
and the Kugo canonical formulation \cite{Kugo}. This section stablishes our
notation, and sets the standard for the new developments that follow.
Section 3 contains the main results, including the construction of the 2PIEA
and the proof of its structure as the sum of 2PI Feynman graphs. In Section
4, we use the 2PIEA as starting point for the discussion of the Zinn -
Justin identity. In Sections 5 and 6 we use the ZJ identiry to study the
gauge dependence and the structure of the propagators, respectively.

We have gathered in the Appendix some relevant formulae concerning Grassmann
calculus. For more details, we refer the reader to the monographs by Berezin 
\cite{Berezin}, DeWitt \cite{DeWitt2} and Negele and Orland \cite{NO98} .

\section{Path Integral approach}

\subsection{Gauge theories}

A gauge theory contains ``matter'' fields $\psi $ such that there are local
(unitary) transformations $g$ which are symmetries of the theory. The $g$'s
form a non abelian (simple) group. Infinitesimal transformations may be
written as $g=exp\left[ i\varepsilon \right] $, where the hermitian matrix $%
\varepsilon $ may be expanded as a linear combination of ``generators'' $%
\varepsilon =\varepsilon ^{A}T_{A}.$ The generators form a closed algebra
under commutation

\begin{equation}
\left[ T_{A},T_{B}\right] =iC_{\;AB}^{C}T_{C}
\end{equation}
The structure constants $C_{\;AB}^{C}$ are antisymmetric on $A,B$ and
satisfy the Jacobi identity.

Gauge invariance of kinetic terms within the Lagrangian means that
derivatives are written in terms of the gauge covariant derivative operator $%
D_{\mu }=\partial _{\mu }-iA_{\mu }$. The conexion $A_{\mu }=A_{\mu A}T^{A}$
transforms upon an infinitesimal gauge transformation as

\begin{equation}
A_{\mu }\rightarrow A_{\mu }+D_{\mu }\varepsilon 
\end{equation}
where

\begin{equation}
D_{\mu }\varepsilon =\partial _{\mu }\varepsilon -i\left[ A_{\mu
},\varepsilon \right]
\end{equation}

Covariant derivatives do not commute, but their commutator contains no
derivatives

\begin{equation}
\left[ D_{\mu },D_{\nu }\right] =-iF_{\mu \nu }
\end{equation}
where the field tensor

\begin{equation}
F_{\mu \nu }=\partial _{\mu }A_{\nu }-\partial _{\nu }A_{\mu }-i\left[
A_{\mu },A_{\nu }\right]
\end{equation}

Upon a gauge transformation

\begin{equation}
F_{\mu \nu }\rightarrow F_{\mu \nu }+i\left[ \varepsilon ,F_{\mu \nu
}\right] 
\end{equation}
therefore the object

\begin{equation}
\frac{-1}{4g^{2}}\mathrm{Tr\,}F^{\mu \nu }F_{\mu \nu }
\end{equation}
is gauge invariant. This is the classical Lagrangian density for the gauge
fields, $g$ being the coupling constant. The total action $S=S_{0}+S_{m}$,
where

\begin{equation}
S_{0}=\int d^{d}x\;\left( \frac{-1}{4g^{2}}\right) \mathrm{Tr\,}F^{\mu \nu
}F_{\mu \nu }
\end{equation}
and $S_{m}$ is the gauge invariant action for the matter fields.

\subsection{DeWitt$^{\prime }$s notation}

We may drop the distinction between gauge and matter fields, and consider a
theory described by a string of fields $\phi ^{\alpha }$ invariant under
infinitesimal transformations

\begin{equation}
\delta \phi ^{\alpha }=T_{A}^{\alpha }\left[ \phi \right] \varepsilon ^{A}
\end{equation}

The commutation rules are the statement that the commutator of two gauge
transforms is also a gauge transform, namely

\begin{equation}
\frac{\delta T_{A}^{\alpha }\left[ \phi \right] }{\delta \phi ^{\beta }}%
T_{B}^{\beta }\left[ \phi \right] -\frac{\delta T_{B}^{\alpha }\left[ \phi
\right] }{\delta \phi ^{\beta }}T_{A}^{\beta }\left[ \phi \right]
=T_{C}^{\alpha }\left[ \phi \right] C_{\;AB}^{C}
\end{equation}

The classical equations of motion read

\begin{equation}
\frac{\delta S}{\delta \phi ^{\alpha }}=0
\end{equation}
and because of gauge invariance we must have the identity

\begin{equation}
\frac{\delta S}{\delta \phi ^{\alpha }}T_{A}^{\alpha }\left[ \phi \right] =0
\end{equation}

\subsection{The vacuum to vacuum amplitude}

In the quantum theory, we expect the vacuum to vacuum amplitude to be given
by the IN-OUT path integral

\begin{equation}
Z=\int D\phi \;e^{iS}
\end{equation}
However this integral counts each history many times, and is ill defined.

To cure this problem, let $f^{A}$ be functionals in history space which are
not gauge invariant. This means that, given a history $\phi ^{\alpha }$ such
that $f^{A}\left[ \phi ^{\alpha }\right] =0,$ and an infinitesimal gauge
transform such that $f^{A}\left[ \phi ^{\alpha }+\delta \phi ^{\alpha
}\right] =0$ too, the gauge transform must be trivial; in other words

\begin{equation}
\mathrm{Det}\,\left[ \frac{\delta f^{A}}{\delta \phi ^{\alpha }}%
T_{B}^{\alpha }\left[ \phi \right] \right] \neq 0
\end{equation}

Now let us call $\phi \left[ \varepsilon \right] $ the result of applying a
gauge transform parameterized by $\varepsilon $ to the field configuration $%
\phi .$ Then we have the identity

\begin{equation}
\int D\varepsilon \;\mathrm{Det}\,\left[ \frac{\delta f^{A}}{\delta \phi
^{\alpha }}\left[ \phi \left[ \varepsilon \right] \right] T_{B}^{\alpha
}\left[ \phi \left[ \varepsilon \right] \right] \right] \;\delta \left[
f^{A}\left[ \phi \left[ \varepsilon \right] \right] -C^{A}\right] =1
\end{equation}
where $C^{A}$ may be anything, and we can write

\begin{equation}
Z=\int D\varepsilon \;\int D\phi \;\mathrm{Det}\,\left[ \frac{\delta f^{A}}{%
\delta \phi ^{\alpha }}\left[ \phi \left[ \varepsilon \right] \right]
T_{B}^{\alpha }\left[ \phi \left[ \varepsilon \right] \right] \right]
\;\delta \left[ f^{A}\left[ \phi \left[ \varepsilon \right] \right]
-C^{A}\right] \;e^{iS\left[ \phi \right] }
\end{equation}
Of course, $S\left[ \phi \right] =S\left[ \phi \left[ \varepsilon \right]
\right] ,$and

\begin{equation}
D\phi \left[ \varepsilon \right] =D\phi \;\left\{ 1+\varepsilon ^{A}\mathrm{%
Tr}\frac{\delta T_{A}^{\alpha }\left[ \phi \right] }{\delta \phi ^{\beta }}%
\right\} 
\end{equation}
so, provided

\begin{equation}
\mathrm{Tr}\frac{\delta T_{A}^{\alpha }\left[ \phi \right] }{\delta \phi
^{\beta }}=0
\end{equation}
we find, up to a constant

\begin{equation}
Z=\int D\phi \;\mathrm{Det}\,\left[ \frac{\delta f^{A}}{\delta \phi ^{\alpha
}}\left[ \phi \right] T_{B}^{\alpha }\left[ \phi \right] \right] \;\delta
\left[ f^{A}\left[ \phi \right] -C^{A}\right] \;e^{iS\left[ \phi \right] }
\end{equation}

Since the $C^{A}$ are arbitrary, any average over different choices will do
too. For example, given a suitable metric we may take the Gaussian average

\begin{equation}
\int DC^{A}\;e^{-\left( i/2\xi \right) C^{A}C_{A}}
\end{equation}
Integrating over $\xi $ and after a Fourier transform we find

\begin{equation}
Z=\int D\phi Dh_{A}\;\mathrm{Det}\,\left[ \frac{\delta f^{A}}{\delta \phi
^{\alpha }}\left[ \phi \right] T_{B}^{\alpha }\left[ \phi \right] \right] \;%
\mathrm{exp}\left\{ i\left[ S\left[ \phi \right] +h_{A}f^{A}\left[ \phi
\right] +\frac{\xi }{2}h^{A}h_{A}\right] \right\} 
\end{equation}
$h_{A}$ is the Nakanishi - Lautrup (N-L) field \cite{Nakanishi}, and $\xi $
the gauge fixing parameter.

\subsection{Ghosts}

We may write the determinant as a functional integral

\begin{equation}
Z=\int D\omega ^{B}D\chi _{A}D\phi Dh_{A}\;\mathrm{exp}\left\{ i\left[
S\left[ \phi \right] +h_{A}f^{A}\left[ \phi \right] +\frac{\xi }{2}%
h^{A}h_{A}+i\chi _{A}\Delta ^{A}\right] \right\}
\end{equation}

\begin{equation}
\Delta ^{A}=\frac{\delta f^{A}}{\delta \phi ^{\alpha }}\left[ \phi \right]
T_{B}^{\alpha }\left[ \phi \right] \omega ^{B}
\end{equation}

The $\omega ^{B}$, $\chi _{A}$ are \textit{independent }c-number Grassmann
variables, namely the ghost and antighost fields, respectively. Following
Kugo and Ojima \cite{Kugo}, and unlike Weinberg \cite{Weinberg2}, we have
included a factor of $i$ in the ghost Lagrangian, which is consistent with
taking the ghosts as formally ``Hermitian'' and demanding the action to be
``real''.

We assign``ghost number'' $1$ to $\omega ^{B}$, and $-1$ to $\chi _{A}.$

\subsection{BRST invariance}

We may regard the functional

\begin{equation}
S_{eff}=S\left[ \phi \right] +h_{A}f^{A}\left[ \phi \right] +\frac{\xi }{2}%
h^{A}h_{A}+i\chi _{A}\Delta ^{A}
\end{equation}
as the action of a new theory, built from the original by adding the N-L,
ghost and antighost fields. By construction, this action is not gauge
invariant in the original sense. However, let us consider a gauge transform
parameterized by $\theta \omega ^{B}$, where $\theta $ is an anticommuting
``constant'', namely

\begin{equation}
\delta \phi ^{\alpha }=\theta T_{A}^{\alpha }\left[ \phi \right] \omega ^{A}
\end{equation}

Observe that, keeping the other fields invariant for the time being

\begin{equation}
\delta f^A\left[ \phi \right] =\theta \Delta ^A
\end{equation}

\begin{equation}
\delta \Delta ^{A}=\theta f_{,\alpha \beta }^{A}T_{B}^{\alpha }\left[ \phi
\right] T_{C}^{\beta }\left[ \phi \right] \omega ^{B}\omega ^{C}+\theta
f_{,\alpha }^{A}T_{B}^{\alpha }\left[ \phi \right] _{,\beta }T_{C}^{\beta
}\left[ \phi \right] \omega ^{C}\omega ^{B}
\end{equation}

Since the ghosts are Grassmann, this becomes

\begin{equation}
\delta \Delta ^{A}=\frac{1}{2}\theta f_{,\alpha }^{A}T_{D}^{\alpha }\left[
\phi \right] C_{\;BC}^{D}\omega ^{C}\omega ^{B}
\end{equation}

These results suggest extending the definition of the transformation to

\begin{equation}
\delta h_A=0
\end{equation}
\begin{equation}
\delta \chi _A=i\theta h_A
\end{equation}

\begin{equation}
\delta \omega ^{D}=\frac{1}{2}\theta C_{\;CB}^{D}\omega ^{C}\omega ^{B}
\end{equation}
Then $S_{eff}$ is invariant under this``BRST'' transformation. Let us define
the operator $\Omega $

\begin{equation}
\Omega \left[ X\right] =\frac{d}{d\theta }\delta X
\end{equation}
The operator $\Omega $ increases``ghost number'' by one. Obviously $\Omega
^{2}\left[ h_{A}\right] =\Omega ^{2}\left[ \chi _{A}\right] =0$, but also

\begin{equation}
\Omega ^{2}\left[ \phi ^{\alpha }\right] =T_{A}^{\alpha }\left[ \phi \right]
_{,\beta }T_{B}^{\beta }\left[ \phi \right] \omega ^{B}\omega ^{A}+\frac{1}{2%
}T_{A}^{\alpha }\left[ \phi \right] C_{\;CB}^{A}\omega ^{C}\omega ^{B}=0
\end{equation}

\begin{equation}
\Omega ^2\left[ \omega ^D\right] =\frac 14C_{\;CB}^D\left[ C_{\;EF}^C\omega
^E\omega ^F\omega ^B+C_{\;EF}^B\omega ^C\omega ^E\omega ^F\right] =0
\end{equation}

These properties imply that actually $\Omega ^{2}=0$ \textit{tout court}
(see \cite{Weinberg2}). Also, observe that

\begin{equation}
S_{eff}=S_{0}+\Omega \left[ F\right]   \label{GFF1}
\end{equation}
where

\begin{equation}
S_{0}=S\left[ \phi \right]   \label{GFF2}
\end{equation}
is BRST invariant, and

\begin{equation}
F=-i\chi _{A}\left\{ f^{A}\left[ \phi \right] +\frac{1}{2}\xi h^{A}\right\} 
\label{GFF3}
\end{equation}
(recall that $\Omega \left[ F\right] =-i\left( \Omega \left[ \chi
_{A}\right] \left\{ f^{A}\left[ \phi \right] +\frac{1}{2}\xi h^{A}\right\}
-\chi _{A}\Omega \left[ f^{A}\left[ \phi \right] \right] \right) $). Also,
observe that, \textit{provided}

\begin{equation}
C_{\;AB}^{A}\equiv 0
\end{equation}
the functional volume element is also BRST invariant.

\subsection{Gauge invariance - gauge independence of the vacuum to vacuum
amplitude}

It follows from the above that any gauge fixing dependence (that is,
dependence on the choice of the gauge fixing condition $f^{A}$, gauge fixing
parameter $\xi $ or the metric used to raise indexes in the N-L field) may
only come from a dependence upon changes in the functional $F$. Any such
change induces a perturbation

\begin{equation}
\delta Z=i\int D\omega ^{B}D\chi _{A}D\phi Dh_{A}\;\Omega \left[ \delta
F\right] \mathrm{exp}\left\{ iS_{eff}\right\}
\end{equation}

Now, call $X^{r}$ the different fields in the theory. Then

\begin{equation}
\Omega \left[ \delta F\right] =\left( -1\right) ^{g_{r}+1}\delta
F_{,r}\Omega \left[ X^{r}\right] 
\end{equation}
where $g_{r}$ is the corresponding ghost number. Integrating by parts (see 
\cite{Gomis}), and \textit{provided the surface term vanishes}, we get

\begin{equation}
\delta Z=i\int D\omega ^{B}D\chi _{A}D\phi Dh_{A}\;\delta F\left\{ \frac{%
\delta }{\delta X^{r}}\mathrm{exp}\left\{ iS_{eff}\right\} \Omega \left[
X^{r}\right] \right\} 
\end{equation}
But the bracketts vanish, because of BRST invariance of $S_{eff}$ and
because $\Omega \left[ X^{r}\right] $ is divergence - free. Therefore the 
\textit{physicality condition} is that the flux of any vector pointing in
the direction of $\Omega \left[ X^{r}\right] $ over the boundary of the
space of field configurations must vanish.

This shows by the way that $F$ could be any expression of ghost number $-1$,
since $S_{eff}$ must have ghost number zero.

\subsection{In-out, vacuum effective action}

Let us write the generating functional for connected Feynman graphs

\begin{equation}
\mathrm{exp}\left\{ iW\right\} =\int D\omega ^{B}D\chi _{A}D\phi Dh_{A}\;%
\mathrm{exp}\left\{ i\left[ S_{eff}+J_{r}X^{r}\right] \right\} 
\end{equation}
and define the in-out effective action as its Legendre transform

\begin{equation}
\Gamma \left[ X^{\alpha }\right] =W\left[ J\right] -J_{r}X^{r}
\end{equation}

Performing a change of variables within the path integral corresponding to a
BRST\ transformation, the generating functional cannot change, and therefore

\begin{equation}
\int D\omega ^{B}D\chi _{A}D\phi Dh_{A}\;\mathrm{exp}\left\{ i\left[
S_{eff}+J_{r}X^{r}\right] \right\} J_{r}\Omega \left[ X^{r}\right] =0
\end{equation}
but

\begin{equation}
J_{r}=-\Gamma _{,r}
\end{equation}
and so we obtain the Zinn - Justin equation

\begin{equation}
\Gamma _{,r}\left\langle \Omega \left[ X^{r}\right] \right\rangle =0
\end{equation}
where

\begin{equation}
\left\langle R\right\rangle =\mathrm{exp}\left\{ -iW\right\} \int D\omega
^{B}D\chi _{A}D\phi Dh_{A}\;\mathrm{exp}\left\{ i\left[
S_{eff}+J_{r}X^{r}\right] \right\} R
\end{equation}

On the other hand, if we simply change the functional $F$ in $S_{eff}$ by an
amount $\delta F$, we get, holding the sources constant

\begin{equation}
\delta W=\mathrm{exp}\left\{ -iW\right\} \int D\omega ^{B}D\chi _{A}D\phi
Dh_{A}\;\Omega \left[ \delta F\right] \mathrm{exp}\left\{ i\left[
S_{eff}+J_{r}X^{r}\right] \right\} 
\end{equation}
and repeating the previous argument

\begin{equation}
\delta W=i\Gamma _{,r}\left\langle \delta F\;\Omega \left[ X^r\right]
\right\rangle
\end{equation}

Of course, if we hold the background fields constant, the sources will
change by an amount $\delta J_{r}$. However, this extra variation does not
contribute to the Legendre transform, and so

\begin{equation}
\delta \Gamma =i\Gamma _{,r}\left\langle \delta F\;\Omega \left[ X^r\right]
\right\rangle
\end{equation}

\subsection{The canonical approach and non - vacuum states}

In this section, we shall consider the concrete case where $\phi ^{\alpha
}=A_{\mu }^{A}$, $f_{,\alpha }^{A}=\delta _{B}^{A}\partial _{\mu }$ and $%
T_{B}^{\alpha }=\delta _{B}^{A}\partial _{\mu }+C_{CB}^{A}A_{\mu }^{C}$. We
can write $S_{eff}$ explicitly

\begin{equation}
S_{eff}=\int d^{4}x\;\left\{ \frac{-1}{4g^{2}}F^{A\mu \nu }F_{A\mu \nu
}-\partial _{\mu }h_{A}A^{\mu A}+\frac{\xi }{2}h^{A}h_{A}-i\partial _{\mu
}\chi _{A}\left[ \delta _{B}^{A}\partial _{\mu }+iC_{CB}^{A}A_{\mu
}^{C}\right] \omega ^{B}\right\}
\end{equation}

If we take $A_{Aa}$ ($a=1,2,3$), $h_{A},$ $\chi _{A}$ and $\omega ^{A}$ as
canonical variables, then we may identify the corresponding momenta \cite
{Kugo}

\begin{equation}
p_{\phi }^{Aa}=\frac{1}{g^{2}}F^{Aa0};\qquad p_{h}^{A}=-A^{A0};\qquad
p_{\chi }^{A}=-i\left[ \delta _{B}^{A}\partial
_{0}+C_{CB}^{A}A_{0}^{C}\right] \omega ^{B};\qquad p_{\omega A}=i\partial
_{0}\chi _{A}
\end{equation}
and impose the ETCCRs

\begin{equation}
\left[ p_{Xr},X^{s}\right] _{\mp }=-i\delta _{s}^{r}
\end{equation}
where we use anticommutators for ghost fields and momenta, and commutators
for all other cases.

The BRST invariance of $S_{eff}$ implies the conservation of the Noether
current

\begin{equation}
j^{\mu }=\Omega \left[ X^{r}\right] \frac{\delta L_{eff}}{\delta \partial
_{\mu }X^{r}}
\end{equation}

We define the BRST charge

\begin{equation}
\Omega =\int d^{3}x\;\Omega \left[ X^{r}\right] p_{Xr}
\end{equation}

This is the generator of BRST transforms, since

\begin{equation}
\delta X^{r}=\theta \Omega \left[ X^{r}\right] =i\left[ \theta \Omega
,X^{r}\right] 
\end{equation}
(since $\theta $ is Grassman, we use commutators throughout). Then $\Omega
^{2}=0.$

$S_{eff}$ is also invariant upon the scale transformation

\begin{equation}
\omega ^{B}\rightarrow e^{\lambda }\omega ^{B},\qquad \chi ^{B}\rightarrow
e^{-\lambda }\chi ^{B}
\end{equation}

The corresponding generator

\begin{equation}
Q=\int d^{3}x\;\left\{ \omega ^{B}p_{\omega B}-\chi _{A}p_{\chi
}^{A}\right\} 
\end{equation}
is the \textit{ghost charge}$.$ Ghost charge is bosonic, so $\left[
Q,Q\right] =0$. On the other hand, $\Omega $ has ghost charge $1$, so

\begin{equation}
i\left[ Q,\Omega \right] =\Omega 
\end{equation}
Both $Q$ and $\theta \Omega $ commute with the effective Hamiltonian.

Observables are BRST invariant, and so they commute with $\theta \Omega .$
Physical states are also BRST invariant, therefore annihilated by $\Omega .$
Physical states differing by a BRST transform are physically
indistinguishable, in the sense that they lead to the same matrix elements
for all observables, and so we may add the condition that a physical state $%
\left| \alpha \right\rangle $ is BRST-closed ($\Omega \left| \alpha
\right\rangle =0$) but not exact (there is no $\left| \beta \right\rangle $
such that $\left| \alpha \right\rangle =\Omega \left| \beta \right\rangle $).

One - particle unphysical states come in tetrads \cite{Kugo}. Let $\left|
N\right\rangle $ be a one - particle unphysical state with ghost charge $N$,
i. e., $iQ\left| N\right\rangle =N\left| N\right\rangle $ (any further BRST
invariant quantum numbers are irrelevant to the argument, and shall be
omitted). Call $\left| N+1\right\rangle =\Omega \left| N\right\rangle .$
Then $\Omega \left| N+1\right\rangle =0$, and this implies that $\left|
N+1\right\rangle $ has vanishing norm, since $\left\langle N+1\right. \left|
N+1\right\rangle =\left\langle N+1\right| \Omega \left| N\right\rangle =0.$
Let $\left| -N-1\right\rangle $ be the state ``conjugated'' to $\left|
N+1\right\rangle $, in the sense that $\left\langle -N-1\right. \left|
N+1\right\rangle =1$ (since $iQ$ is hermitian, this state must have ghost
charge $-N-1$). Now $\left\langle -N-1\right| \Omega \left| N\right\rangle
=1 $, and so $\left| -N\right\rangle =\Omega \left| -N-1\right\rangle $ is
conjugated to $\left| N\right\rangle .$ Observe that, without loss of
generality, we may assume that $N$ is even.

Now let $a_{M}^{\dagger }$ be the corresponding creation operators, namely $%
\left| M\right\rangle =a_{M}^{\dagger }\left| 0\right\rangle $ ($M=N,$ $-N,$ 
$N+1$ or $-N-1$). Then $a_{N+1}^{\dagger }=\left[ \Omega ,a_{N}^{\dagger
}\right] $ and $a_{-N}^{\dagger }=\left\{ \Omega ,a_{-N-1}^{\dagger
}\right\} $. Now $\Omega ^{2}=0$, so $\left\{ \Omega ,a_{N+1}^{\dagger
}\right\} =\left[ \Omega ,a_{-N}^{\dagger }\right] =0.$ Because $\left|
N+1\right\rangle $ and $\left| -N\right\rangle $ have vanishing norms, we
must have $\left\{ a_{N+1},a_{N+1}^{\dagger }\right\} =\left[
a_{-N},a_{-N}^{\dagger }\right] =0$; the conjugacy relations imply $\left\{
a_{-N-1},a_{N+1}^{\dagger }\right\} =\left[ a_{-N},a_{N}^{\dagger }\right]
=1.$ The $a_{M}$'s are destruction operators, but $a_{N}$ destroys the $%
\left| -N\right\rangle $ (rather than the $\left| N\right\rangle $) state, $%
a_{N+1}$ destroys $\left| -N-1\right\rangle ,$ etc.

It follows that the projection operator $P_{1}$ over the subspace of states
with exactly one unphysical particle may be written in terms of the
projector $P$ over physical states as

\begin{equation}
P_{1}=a_{-N}^{\dagger }Pa_{N}+a_{N}^{\dagger }Pa_{-N}+a_{-N-1}^{\dagger
}Pa_{N+1}+a_{N+1}^{\dagger }Pa_{-N-1}
\end{equation}
But then

\begin{eqnarray}
P_{1} &=&\Omega \left[ a_{-N-1}^{\dagger }Pa_{N}+a_{N}^{\dagger
}Pa_{-N-1}\right]  \nonumber \\
&&+a_{-N-1}^{\dagger }\Omega Pa_{N}+a_{N}^{\dagger
}Pa_{-N}+a_{-N-1}^{\dagger }Pa_{N+1}-a_{N}^{\dagger }\Omega Pa_{-N-1} 
\nonumber \\
&=&\Omega \left[ a_{-N-1}^{\dagger }Pa_{N}+a_{N}^{\dagger }Pa_{-N-1}\right] 
\nonumber \\
&&+a_{-N-1}^{\dagger }P\Omega a_{N}+a_{N}^{\dagger
}Pa_{-N}+a_{-N-1}^{\dagger }Pa_{N+1}-a_{N}^{\dagger }P\Omega a_{-N-1} 
\nonumber \\
&=&\Omega \left[ a_{-N-1}^{\dagger }Pa_{N}+a_{N}^{\dagger }Pa_{-N-1}\right] 
\nonumber \\
&&+a_{-N-1}^{\dagger }P\Omega a_{N}+a_{N}^{\dagger }P\left\{ \Omega
,a_{-N-1}\right\} -a_{-N-1}^{\dagger }P\left[ \Omega ,a_{N}\right]
-a_{N}^{\dagger }P\Omega a_{-N-1}  \nonumber \\
&=&\left\{ \Omega ,a_{-N-1}^{\dagger }Pa_{N}+a_{N}^{\dagger
}Pa_{-N-1}\right\}
\end{eqnarray}

Repeating identical arguments for the spaces with $n$ unphysical particles,
we conclude that the projector $P^{\prime }$ orthogonal to $P$ has the form $%
P^{\prime }=\left\{ \Omega ,R\right\} $, where $R$ is some operator with
ghost number $-1$.

We may now deal with the construction of statistical operators in gauge
theories. In principle, a physical statistical operator should shield
nonzero probabilities only for physical states, and so it should satisfy $%
\rho =P\rho =\rho P.$ This is a much stronger requirement than BRST
invariance $\left[ \Omega ,\rho \right] =0.$ So, given a BRST invariant
density matrix $\rho $, we ought to define the physical expectation value of
any (BRST invariant) observable $C$ as

\begin{equation}
\left\langle C\right\rangle _{phys}=\mathrm{Tr}\left[ P\rho C\right]
\end{equation}

However, Kugo and Hata \cite{Kugo} (KH) have shown that the same expectation
values may be obtained by using the statistical operator $e^{-\pi Q}\rho $ .
The key to the argument is that the commutation relation $\left[ iQ,\Omega
\right] =\Omega $ implies that, if $\left| N\right\rangle $ is an eigenstate
of $iQ$ with eigenvalue $N$, then $\Omega \left| N\right\rangle $ has
eigenvalue $N+1$. It follows that $\left\{ e^{-\pi Q},\Omega \right\} =0,$
since $e^{-\pi Q}=e^{i\pi \left( iQ\right) }$. We then find that, for any
BRST invariant observable $C$

\begin{equation}
\left\langle C\right\rangle _{phys}=\mathrm{Tr}\left[ P\rho C\right] =%
\mathrm{Tr}\left[ Pe^{-\pi Q}\rho C\right] =\mathrm{Tr}\left[ e^{-\pi Q}\rho
C\right] -\mathrm{Tr}\left[ \left\{ \Omega ,R\right\} e^{-\pi Q}\rho
C\right] 
\end{equation}
We must show that the second term vanishes, and this follows from $\left\{
e^{-\pi Q},\Omega \right\} =0$ and $\left[ \Omega ,\rho C\right] =0$.

This suggest to define the expectation value $\left\langle C\right\rangle $
of any observable as $\left\langle C\right\rangle =\mathrm{Tr}\left[ e^{-\pi
Q}\rho C\right] $. Of course, this agrees with the physical expectation
value only if $C$ is BRST invariant. For example, the partition function
computed from $e^{-\pi Q}\rho $ agrees with the partition function defined
by tracing only over physical states, but the generating functionals
obtained by adding sources coupled to non - BRST invariant operators will in
general be different.

The advantages of the Kugo - Hata ansatz are clearly seen by considering the
form of the KMS theorem appropiate to the ghost propagator. Let us define

\begin{equation}
G_{AB}^{ab}\left( x,x^{\prime }\right) =\left\langle P\left[ \chi
_{A}^{a}\left( x\right) \omega _{B}^{b}\left( x^{\prime }\right) \right]
\right\rangle 
\end{equation}
where $P$ is the usual (CTP)-ordering operator. Then

\begin{equation}
G_{AB}^{21}\left( x,x^{\prime }\right) =\left\langle \chi _{A}\left(
x\right) \omega _{B}\left( x^{\prime }\right) \right\rangle
\end{equation}

\begin{equation}
G_{AB}^{12}\left( x,x^{\prime }\right) =-\left\langle \omega _{B}\left(
x^{\prime }\right) \chi _{A}\left( x\right) \right\rangle 
\end{equation}
(observe the sign change, associated to the anticommuting character of the
ghost fields). The Jordan propagator is defined as $G=G^{21}-G^{12}$.

If we omitted the K-H $e^{-\pi Q}$ factor, we would reason, given $\rho
=e^{-\beta H}$,

\begin{equation}
G_{AB}^{21}\left( x,x^{\prime }\right) \approx \mathrm{Tr}\left[ e^{-\beta
H}\chi _{A}\left( x\right) \omega _{B}\left( x^{\prime }\right) \right] =%
\mathrm{Tr}\left[ \chi _{A}\left( x+i\beta \right) e^{-\beta H}\omega
_{B}\left( x^{\prime }\right) \right] =-G_{AB}^{12}\left( x+i\beta
,x^{\prime }\right) 
\end{equation}
Therefore $G_{AB}^{21}\left( \omega \right) =-e^{\beta \omega
}G_{AB}^{12}\left( \omega \right) ,$ leading to a Fermi - Dirac form of the
thermal propagators.

This reasoning is incorrect. The proper way is

\begin{equation}
G_{AB}^{21}\left( x,x^{\prime }\right) =\mathrm{Tr}\left[ e^{-\pi Q}\chi
_{A}\left( x+i\beta \right) e^{-\beta H}\omega _{B}\left( x^{\prime }\right)
\right] =G_{AB}^{12}\left( x+i\beta ,x^{\prime }\right)
\end{equation}

So $G_{AB}^{21}\left( \omega \right) =e^{\beta \omega }G_{AB}^{12}\left(
\omega \right) $, which leads to the Bose - Einstein form.

Let us observe that in the path integral representation, the K-H factor does
not appear explicitly, but only changes the boundary conditions on ghost
fields from anti-periodic to periodic.

We conclude that in this formalism, unphysical degrees of freedom and ghosts
get thermal corrections, both being of the Bose - Einstein form, in spite of
the ghosts being fermions (for which reason ghost loops do get a minus
sign). For an alternative formulation, see \cite{Landshoff}

\section{The 2PI formalism applied to gauge theories}

We can now begin with our real goal, namely, the application of the 2PI CTP
formalism to gauge theories. We shall proceed with a fair amount of
generality, only assuming that the gauge condition is linear, and also the
gauge generators $T_{A}^{\alpha }\left[ \phi \right] =T_{0A}^{\alpha
}+T_{1A\beta }^{\alpha }\phi ^{\beta }$.

The classical action is given by

\begin{equation}
S_{eff}=S\left[ \phi \right] +h_{A}f^{A}\left[ \phi \right] +\frac{\xi }{2}%
h^{A}h_{A}+i\chi _{A}\frac{\delta f^{A}}{\delta \phi ^{\alpha }}\left[ \phi
\right] T_{B}^{\alpha }\left[ \phi \right] \omega ^{B}  \label{CLAS}
\end{equation}
To this we add sources coupled to the individual degrees of freedom and also
to their products

\begin{equation}
X^{r}J_{r}+\frac{1}{2}X^{r}\mathbf{K}_{rs}X^{s}=j_{\alpha }x^{\alpha
}+\theta ^{u}\lambda _{u}+\frac{1}{2}\kappa _{\alpha \beta }x^{\alpha
}x^{\beta }+\frac{1}{2}\sigma _{uv}\theta ^{u}\theta ^{v}+\theta ^{u}\psi
_{u\alpha }x^{\alpha }
\end{equation}
where $x^{\alpha }$ represents the bosonic degrees of freedom ($\phi $, $h$)
and $\theta $ the Grassmann ones ($\omega $, $\chi $), and we introduce the
definition $\mathbf{K}_{\alpha u}=-\mathbf{K}_{u\alpha }$. Observe that $j$, 
$\kappa $ and $\sigma $ are normal, while $\lambda $ and $\psi $ are
Grassmann. $\sigma $ is antisymmetric.

We therefore define the generating functional

\begin{equation}
e^{iW}=\int DX^r\;exp\left\{ i\left[ S_{eff}+X^rJ_r+\frac 12X^r\mathbf{K}%
_{rs}X^s\right] \right\}
\end{equation}

Note that the information about the initial state is implicit in the
integration measure, and will reappear only as an initial condition on the
equations of motion.

We find

\begin{equation}
W\overleftarrow{\frac \delta {\delta J_r}}=\bar X^r
\end{equation}

\begin{equation}
W\overleftarrow{\frac{\delta }{\delta \mathbf{K}_{rs}}}=\frac{\theta
^{sr}\theta ^{s}}{2}\left[ \bar{X}^{r}\bar{X}^{s}+\mathbf{G}^{rs}\right] 
\end{equation}
where we introduce the bookkeeping devise $\theta ^{r}=\left( -1\right)
^{q_{r}}$, where $q_{r}$ is the ghost charge of the corresponding field, and 
$\theta ^{rs}=\left( -1\right) ^{q_{r}q_{s}}$.

We define the Legendre transform

\begin{equation}
\Gamma =W-\bar{X}^{r}J_{r}-\frac{1}{2}\bar{X}^{r}\mathbf{K}_{rs}\bar{X}^{s}-%
\frac{\theta ^{sr}\theta ^{s}}{2}\mathbf{G}^{rs}\mathbf{K}_{rs}
\end{equation}
whereby

\begin{eqnarray}
\frac{\delta }{\delta X^{r}}\Gamma &=&\theta ^{rs}\left[ W\overleftarrow{%
\frac{\delta }{\delta J_{s}}}\right] \left[ \frac{\delta }{\delta X^{r}}%
J_{s}\right] +\theta ^{rs}\theta ^{rt}\left[ W\overleftarrow{\frac{\delta }{%
\delta \mathbf{K}_{st}}}\right] \left[ \frac{\delta }{\delta X^{r}}\mathbf{K}%
_{st}\right]  \nonumber \\
&&\ -J_{r}-\theta ^{rs}\bar{X}^{s}\left[ \frac{\delta }{\delta X^{r}}%
J_{s}\right] -\frac{1}{2}\mathbf{K}_{rs}\bar{X}^{s}-\frac{1}{2}\theta ^{rs}%
\bar{X}^{s}\left[ \frac{\delta }{\delta X^{r}}\mathbf{K}_{st}\right] \bar{X}%
^{t}  \nonumber \\
&&-\frac{1}{2}\theta ^{sr}\bar{X}^{s}\mathbf{K}_{sr}-\frac{\theta
^{st}\theta ^{t}}{2}\theta ^{rs}\theta ^{rt}\mathbf{G}^{st}\left[ \frac{%
\delta }{\delta X^{r}}\mathbf{K}_{st}\right]  \nonumber \\
\ &=&\theta ^{rs}\theta ^{rt}\frac{\theta ^{st}\theta ^{t}}{2}\bar{X}^{s}%
\bar{X}^{t}\left[ \frac{\delta }{\delta X^{r}}\mathbf{K}_{st}\right] -J_{r}-%
\frac{1}{2}\mathbf{K}_{rs}\bar{X}^{s}-\frac{1}{2}\theta ^{rs}\bar{X}%
^{s}\left[ \frac{\delta }{\delta X^{r}}\mathbf{K}_{st}\right] \bar{X}^{t}-%
\frac{1}{2}\theta ^{r}\bar{X}^{s}\mathbf{K}_{sr}  \nonumber \\
\ &=&-J_{r}-\frac{1}{2}\mathbf{K}_{rs}\bar{X}^{s}-\frac{1}{2}\theta ^{r}\bar{%
X}^{s}\mathbf{K}_{sr}
\end{eqnarray}

Now observe that $\mathbf{K}_{sr}=\theta ^{r}\theta ^{s}\theta ^{rs}\mathbf{K%
}_{rs} $. In the end

\begin{equation}
\frac{\delta }{\delta X^{r}}\Gamma =-J_{r}-\mathbf{K}_{rs}\bar{X}^{s}
\end{equation}

In the same way

\begin{eqnarray}
\frac{\delta }{\delta \mathbf{G}^{rs}}\Gamma &=&\theta ^{rt}\theta
^{st}\left[ W\overleftarrow{\frac{\delta }{\delta J_{t}}}\right] \left[ 
\frac{\delta }{\delta \mathbf{G}^{rs}}J_{t}\right] +\theta ^{rt}\theta
^{st}\theta ^{rq}\theta ^{sq}\left[ W\overleftarrow{\frac{\delta }{\delta 
\mathbf{K}_{tq}}}\right] \left[ \frac{\delta }{\delta \mathbf{G}^{rs}}%
\mathbf{K}_{tq}\right]  \nonumber \\
&&-\theta ^{rt}\theta ^{st}\bar{X}^{t}\left[ \frac{\delta }{\delta \mathbf{G}%
^{rs}}J_{t}\right] -\frac{1}{2}\theta ^{rt}\theta ^{st}\bar{X}^{t}\left[ 
\frac{\delta }{\delta \mathbf{G}^{rs}}\mathbf{K}_{tq}\right] \bar{X}^{q} 
\nonumber \\
&&-\frac{\theta ^{sr}\theta ^{s}}{2}\mathbf{K}_{rs}-\frac{\theta ^{tq}\theta
^{q}}{2}\theta ^{rt}\theta ^{st}\theta ^{rq}\theta ^{sq}\mathbf{G}%
^{tq}\left[ \frac{\delta }{\delta \mathbf{G}^{rs}}\mathbf{K}_{tq}\right] 
\nonumber \\
&=&\theta ^{rt}\theta ^{st}\theta ^{rq}\theta ^{sq}\frac{\theta ^{tq}\theta
^{q}}{2}\bar{X}^{t}\bar{X}^{q}\left[ \frac{\delta }{\delta \mathbf{G}^{rs}}%
\mathbf{K}_{tq}\right] -\frac{1}{2}\theta ^{rt}\theta ^{st}\bar{X}^{t}\left[ 
\frac{\delta }{\delta \mathbf{G}^{rs}}\mathbf{K}_{tq}\right] \bar{X}^{q}-%
\frac{\theta ^{sr}\theta ^{s}}{2}\mathbf{K}_{rs}  \nonumber \\
&=&-\frac{\theta ^{sr}\theta ^{s}}{2}\mathbf{K}_{rs}
\end{eqnarray}

\subsection{The 2PIEA}

In order to evaluate the 2PIEA, let us make the ansatz

\begin{equation}
\Gamma =\bar{S}\left[ \bar{X}^{r}\right] +\frac{1}{2}\theta ^{sr}\theta ^{s}%
\mathbf{G}^{rs}\mathbf{S}_{rs}-\frac{i}{2}\ln \mathrm{sdet}\left[ \mathbf{G}%
^{rs}\right] +\Gamma _{2}-\frac{i}{2}\theta ^{s}\mathbf{G}^{rs}\mathbf{G}%
_{Rsr}^{-1};  \label{TPIEA}
\end{equation}
where

\begin{equation}
\mathbf{S}_{rs}=\left[ \overrightarrow{\frac{\delta }{\delta \bar{X}^{r}}}%
\bar{S}\right] \overleftarrow{\frac{\delta }{\delta \bar{X}^{s}}}
\end{equation}
and $\bar{S}$ is the classical action Eq. (\ref{CLAS}).

The corresponding ansatze for the sources are

\begin{equation}
J_{r}+\mathbf{K}_{rs}\bar{X}^{s}=-\left[ \frac{\delta }{\delta \bar{X}^{r}}%
\Gamma \right] =-\bar{S}_{,r}+J_{2r}
\end{equation}

\begin{equation}
\mathbf{K}_{rs}=-2\theta ^{sr}\theta ^{s}\left[ \frac{\delta }{\delta 
\mathbf{G}^{rs}}\Gamma \right] =-\mathbf{S}_{rs}+i\mathbf{G}_{Lrs}^{-1}+%
\mathbf{K}_{2rs}
\end{equation}

The generating functional

\begin{eqnarray}
W &=&\Gamma +\bar{X}^{r}J_{r}+\frac{1}{2}\bar{X}^{r}\mathbf{K}_{rs}\bar{X}%
^{s}+\frac{\theta ^{sr}\theta ^{s}}{2}\mathbf{G}^{rs}\mathbf{K}_{rs} 
\nonumber \\
\  &=&\bar{S}\left[ \bar{X}^{r}\right] +\frac{1}{2}\theta ^{sr}\theta ^{s}%
\mathbf{G}^{rs}\mathbf{S}_{rs}-\frac{i}{2}\ln \mathrm{sdet}\left[ \mathbf{G}%
^{rs}\right] +\Gamma _{2}-\frac{i}{2}\theta ^{s}\mathbf{G}^{rs}\mathbf{G}%
_{Rsr}^{-1}+\bar{X}^{r}\left[ -\bar{S}_{,r}+J_{2r}\right]   \nonumber \\
&&\ \ \ -\frac{1}{2}\bar{X}^{r}\left[ -\mathbf{S}_{rs}+i\mathbf{G}%
_{Lrs}^{-1}+\mathbf{K}_{2rs}\right] \bar{X}^{s}+\frac{\theta ^{sr}\theta ^{s}%
}{2}\mathbf{G}^{rs}\left[ -\mathbf{S}_{rs}+i\mathbf{G}_{Lrs}^{-1}+\mathbf{K}%
_{2rs}\right]   \nonumber \\
\  &=&\bar{S}\left[ \bar{X}^{r}\right] -\frac{i}{2}\ln \mathrm{sdet}\left[ 
\mathbf{G}^{rs}\right] +\Gamma _{2}-\bar{X}^{r}\left( \bar{S}_{,r}-J_{2r}-%
\frac{1}{2}\left( \mathbf{S}_{rs}-i\mathbf{G}_{Lrs}^{-1}-\mathbf{K}%
_{2rs}\right) \bar{X}^{s}\right)   \nonumber \\
&&+\frac{\theta ^{sr}\theta ^{s}}{2}\mathbf{G}^{rs}\mathbf{K}_{2rs}
\end{eqnarray}
so

\begin{eqnarray}
&&\ \ \ \left( \mathrm{sdet}\left[ \mathbf{G}^{rs}\right] \right) ^{1/2}\exp
\left[ i\Gamma _{2}\right]   \nonumber \\
\  &=&\int DX^{r}\;\exp i\left[ S\left( X\right) +X^{r}\left( -\bar{S}%
_{,r}+J_{2r}\right) \right.   \nonumber \\
&&+\frac{1}{2}X^{r}\left( -\mathbf{S}_{rs}+i\mathbf{G}_{Lrs}^{-1}+\mathbf{K}%
_{2rs}\right) X^{s}-X^{r}\left( -\mathbf{S}_{rs}+i\mathbf{G}_{Lrs}^{-1}+%
\mathbf{K}_{2rs}\right) \bar{X}^{s}  \nonumber \\
&&\ \ \left. -\bar{S}\left[ \bar{X}^{r}\right] +\bar{X}^{r}\left( \bar{S}%
_{,r}-\frac{1}{2}\mathbf{S}_{rs}\bar{X}^{s}+\frac{i}{2}\mathbf{G}_{Lrs}^{-1}%
\bar{X}^{s}-J_{2r}+\frac{1}{2}\mathbf{K}_{2rs}\bar{X}^{s}\right) -\frac{1}{2}%
\mathbf{G}^{rs}\theta ^{sr}\theta ^{s}\mathbf{K}_{2rs}\right] 
\end{eqnarray}
or else

\begin{equation}
e^{i\Gamma _{2}}=\left( \mathrm{sdet}\left[ \mathbf{G}^{rs}\right] \right)
^{-1/2}\int D\delta X^{r}\;\exp i\left[ \Delta S+\frac{i}{2}\delta X^{r}%
\mathbf{G}_{Lrs}^{-1}\delta X^{s}+\delta X^{r}J_{2r}+\frac{\theta
^{sr}\theta ^{s}}{2}\left[ \delta X^{r}\delta X^{s}-\mathbf{G}^{rs}\right] 
\mathbf{K}_{2rs}\right]   \label{gamma2}
\end{equation}
where

\begin{equation}
\Delta S=S\left[ \bar{X}^{r}+\delta X^{r}\right] -\bar{S}\left[ \bar{X}%
^{r}\right] -\delta X^{r}\bar{S}_{,r}-\frac{1}{2}\delta X^{r}\mathbf{S}%
_{rs}\delta X^{s}
\end{equation}
$\Gamma _{2}$ is the sum of 2PI vacuum bubbles in a theory with free action $%
i\mathbf{G}_{Lrs}^{-1}$ and interacting terms coming from the cubic and
quartic terms in the development of $\bar{S}$ around the mean fields.

In spite of appearances, the new term $\theta ^{s}\mathbf{G}^{rs}\mathbf{G}%
_{Rsr}^{-1}$ is a constant. To see this, parametrize

\begin{equation}
\mathbf{G}^{rs}=\left( 
\begin{array}{ll}
H^{\alpha \beta } & N^{v\alpha } \\ 
N^{u\beta } & M^{uv}
\end{array}
\right) 
\end{equation}
leading to

\begin{equation}
\mathbf{G}_{Rsr}^{-1}=\left( 
\begin{array}{ll}
\bar{H}_{\alpha \beta } & \bar{N}_{v\alpha } \\ 
\bar{N}_{u\beta } & \bar{M}_{uv}
\end{array}
\right) 
\end{equation}
Then, because they are inverses, we must have

\[
H^{\alpha \beta }\bar H_{\beta \gamma }+N^{v\alpha }\bar N_{v\gamma }=\delta
_\gamma ^\alpha 
\]

\[
H^{\alpha \beta }\bar N_{v\beta }+N^{u\alpha }\bar M_{uv}=0 
\]

\[
N^{u\beta }\bar H_{\beta \gamma }+M^{uv}\bar N_{v\gamma }=0 
\]

\begin{equation}
N^{u\beta }\bar{N}_{v\beta }+M^{uw}\bar{M}_{wv}=\delta _{v}^{u}
\end{equation}
and therefore

\begin{eqnarray}
\theta ^{s}\mathbf{G}^{rs}\mathbf{G}_{Rsr}^{-1} &=&\left( H^{\alpha \beta }%
\bar{H}_{\beta \alpha }-N^{v\alpha }\bar{N}_{v\alpha }\right) +\left(
N^{u\beta }\bar{N}_{u\beta }-M^{uv}\bar{M}_{vu}\right)   \nonumber \\
&=&\delta _{\alpha }^{\alpha }-2N^{v\alpha }\bar{N}_{v\alpha }-\left( \delta
_{u}^{u}-2N^{u\beta }\bar{N}_{u\beta }\right) =\delta _{\alpha }^{\alpha
}-\delta _{u}^{u}
\end{eqnarray}
independent of $\mathbf{G}^{rs}.$ It may therefore be discarded.

\subsection{The reduced 2PIEA}

Let us now investigate the Schwinger-Dyson equations

\[
\frac \delta {\delta X^r}\Gamma =0 
\]

\begin{equation}
\frac{\delta }{\delta \mathbf{G}^{rs}}\Gamma =0
\end{equation}
From Eq. (\ref{TPIEA}) we get

\[
\frac \delta {\delta X^r}\bar S\left[ \bar X^r\right] +\frac 12\theta
^{pq}\theta ^q\theta ^{rp}\theta ^{rq}\mathbf{G}^{pq}\frac \delta {\delta
X^r}\mathbf{S}_{pq}+\frac \delta {\delta X^r}\Gamma _2=0 
\]

\begin{equation}
\theta ^{sr}\theta ^{s}\mathbf{S}_{rs}-i\theta ^{r}\theta ^{rs}\left( 
\mathbf{G}_{R}^{-1}\right) _{rs}+2\frac{\delta }{\delta \mathbf{G}^{rs}}%
\Gamma _{2}=0
\end{equation}

The second set of equations may be rewritten as

\begin{equation}
\theta ^{sr}\theta ^{s}\mathbf{S}_{rs}-i\theta ^{r}\left( \mathbf{G}%
_{L}^{-1}\right) _{sr}+2\frac{\delta }{\delta \mathbf{G}^{rs}}\Gamma _{2}=0
\end{equation}
and finally as

\begin{equation}
\mathbf{S}_{rs}-i\left( \mathbf{G}_{L}^{-1}\right) _{rs}+2\theta ^{sr}\theta
^{s}\frac{\delta }{\delta \mathbf{G}^{rs}}\Gamma _{2}=0
\end{equation}

The classical action is given by Eq. (\ref{CLAS}). If we expand $X^{r}=\bar{X%
}^{r}+\delta X^{r}$, then the quadratic terms are

\begin{eqnarray}
\bar{S}^{\left( 2\right) } &=&S_{c}^{\left( 2\right) }\left[ \bar{\phi}%
,\delta \phi \right] +\delta h_{A}f_{\alpha }^{A}\delta \phi ^{\alpha }+%
\frac{\xi }{2}\delta h^{A}\delta h_{A}+i\delta \chi _{A}f_{\alpha
}^{A}T_{B}^{\alpha }\left[ \bar{\phi}\right] \delta \omega ^{B}  \nonumber \\
&&+i\bar{\chi}_{A}f_{\alpha }^{A}T_{1B\beta }^{\alpha }\delta \phi ^{\beta
}\delta \omega ^{B}+i\delta \chi _{A}f_{\alpha }^{A}T_{1B\beta }^{\alpha
}\delta \phi ^{\beta }\bar{\omega}^{B}  \label{QUAD}
\end{eqnarray}
The cubic and quartic terms are

\begin{equation}
\bar{S}^{\left( 3+\right) }=S_{c}^{\left( 3+\right) }\left[ \bar{\phi}%
,\delta \phi \right] +i\delta \chi _{A}f_{\alpha }^{A}T_{1B\beta }^{\alpha
}\delta \phi ^{\beta }\delta \omega ^{B}  \label{CUB}
\end{equation}
Observe that $\Gamma _{2}$ is independent of the gauge fixing parameter, and
that there are no $h$ field lines. To take advantage of this fact, it is
convenient \textit{not }to couple sources to the $h$ field. In this way, $%
\Gamma _{2}$ is independent of the $h$ field, and the respective variations
are exact, namely

\[
f_\alpha ^A\bar \phi ^\alpha +\xi \bar h^A=0 
\]

\[
\xi \delta _{AB}-i\left[ \mathbf{G}_L^{-1}\right] _{hhAB}=0 
\]

\[
f_\alpha ^A-i\left[ \mathbf{G}_L^{-1}\right] _{h\phi \alpha }^A=0 
\]

\begin{equation}
\left[ \mathbf{G}_{L}^{-1}\right] _{h\omega B}^{A}=\left[ \mathbf{G}%
_{L}^{-1}\right] _{h\chi B}^{A}=0
\end{equation}

We may then write the equations

\begin{equation}
\left[ \mathbf{G}^{-1}\right] _{h\phi A\beta }G_{\phi X}^{\beta r}+\left[ 
\mathbf{G}^{-1}\right] _{hhAB}G_{hX}^{Br}=\delta _{Xh^{C}}\delta _{A}^{C}
\end{equation}
as

\begin{equation}
f_{\beta }^{A}G_{\phi X}^{\beta r}+\xi G_{hX}^{Ar}=i\delta _{Xh^{C}}\delta
^{AC}
\end{equation}
namely

\begin{equation}
G_{hX}^{Ar}=\frac{-1}{\xi }f_{\beta }^{A}G_{\phi X}^{\beta r},\qquad X=\phi
,\chi ,\omega 
\end{equation}
and

\begin{equation}
G_{hhA}^{C}=\frac{1}{\xi }\left[ -f_{A\beta }G_{\phi h}^{\beta C}+i\delta
_{A}^{C}\right] =\frac{1}{\xi }\left[ \frac{1}{\xi }f_{A\beta }f_{\gamma
}^{C}G_{\phi \phi }^{\beta \gamma }+i\delta _{A}^{C}\right] 
\end{equation}
which is of course what we expect from the N-L field being Gaussian. We
could use these formulae to actually elliminate the N-L field from the
2PIEA, thus obtaining a reduced effective action.

We shall assume that all fields with non zero ghost number vanish, i. e.,

\begin{equation}
\bar{\omega}=\bar{\chi}=G_{\omega \omega }=G_{\chi \chi }=G_{\omega \phi
}=G_{\omega h}=G_{\chi \phi }=G_{\chi h}=0
\end{equation}
Since the effective action itself has zero ghost number, it cannot contain
terms linear on any of the above, and therefore this condition is consistent
with the equations of motion.

Given these conditions, we have, besides the equations determining the $h$
propagators, the further equations

\[
\left[ \mathbf{G}^{-1}\right] _{\phi \phi \alpha \beta }G_{\phi \phi
}^{\beta \gamma }+\left[ \mathbf{G}^{-1}\right] _{\phi h\alpha B}G_{h\phi
}^{B\gamma }=\delta _{\alpha }^{\gamma } 
\]

\begin{equation}
\left[ \mathbf{G}^{-1}\right] _{\phi \phi \alpha \beta }G_{\phi h}^{\beta
C}+\left[ \mathbf{G}^{-1}\right] _{\phi h\alpha B}G_{hh}^{BC}=0
\end{equation}
leading to 
\[
\left[ \left[ \mathbf{G}^{-1}\right] _{\phi \phi \alpha \beta }+\frac{i}{\xi 
}f_{B\alpha }f_{\beta }^{B}\right] G_{\phi \phi }^{\beta \gamma }=\delta
_{\alpha }^{\gamma }
\]

\begin{equation}
\left[ \left[ \mathbf{G}^{-1}\right] _{\phi \phi \alpha \beta }+\frac{i}{\xi 
}f_{B\alpha }f_{\beta }^{B}\right] G_{\phi h}^{\beta C}=\frac{1}{\xi }%
f_{B\alpha }\delta ^{BC}
\end{equation}

The second one gives nothing new, and the first yields

\begin{equation}
\left[ \mathbf{G}^{-1}\right] _{\phi \phi \alpha \beta }=\left[ G_{\phi \phi
}^{-1}\right] _{\alpha \beta }-\frac{i}{\xi }f_{B\alpha }f_{\beta }^{B}
\end{equation}
so finally we get the equation for the gluon propagator

\begin{equation}
S_{c,\alpha \beta }-\frac{1}{\xi }f_{B\alpha }f_{\beta }^{B}-i\left[ G_{\phi
\phi }^{-1}\right] _{\alpha \beta }+2\frac{\delta \Gamma _{2}}{\delta
G_{\phi \phi }^{\alpha \beta }}=0
\end{equation}

The other nontrivial equation is

\begin{equation}
-if_{\alpha }^{A^{\prime }}T_{B}^{\alpha }\left[ \bar{\phi}\right] +i\left[ 
\mathbf{G}_{L}^{-1}\right] _{\omega \chi B}^{A^{\prime }}+2\frac{\delta
\Gamma _{2}}{\delta G_{\chi \omega A}^{B}}=0
\end{equation}

In deriving this equation we must consider $G_{\chi \omega A}^{B}$ and $%
G_{\omega \chi A}^{B}$ as independent quantities.

\subsection{The abelian case}

In the abelian case the classical action is quadratic, the 2PIEA is exact ($%
\Gamma _2=0$). Since we know that $S_{,\alpha \beta }T_B^\alpha =0$, it is
natural to decompose $G^{\alpha \beta }=G_T^{\alpha \beta }+G_L^{\alpha
\beta }$, where $f_{,\alpha }^BG_T^{\alpha \beta }=S_{,\alpha \beta
}G_L^{\beta \gamma }=0$. Then

\begin{equation}
S_{,\alpha \beta }G_{T}^{\beta \gamma }-\frac{1}{\xi }f_{,\alpha
}^{B}f_{B,\beta }G_{L}^{\beta \gamma }=i\delta _{\alpha }^{\gamma }
\end{equation}

Observe that

\begin{equation}
P_{L\beta }^{\alpha }=f_{B,\beta }\left[ f_{B,\gamma }T_{D}^{\gamma }\right]
^{-1}T_{D}^{\alpha }
\end{equation}
satisfies $P_{L}^{2}=P_{L}$ and kills vectors orthogonal to the gauge
direction. In particular, $P_{L\gamma }^{\alpha }S_{,\alpha \beta }=0,$ so

\begin{equation}
f_{,\delta }^{B}f_{B,\beta }G_{L}^{\beta \gamma }=4i\xi P_{L\delta }^{\gamma
}\Rightarrow G_{L}^{\alpha \beta }=4i\xi \left[ f_{,\alpha }^{B}f_{B,\gamma
}\right] ^{-1}P_{L\gamma }^{\beta }
\end{equation}
and

\begin{equation}
S_{,\alpha \beta }G_{T}^{\beta \gamma }=i\left( \delta _{\alpha }^{\gamma
}-P_{L\alpha }^{\gamma }\right)
\end{equation}

This shows that the dispersion relations in the transverse part are
determined by the zeroes of the classical action, and are therefore gauge
independent, unless $\left( \delta _\alpha ^\gamma -P_{L\alpha }^\gamma
\right) $ is pathological.

\subsection{Some explicit formulae}

As before, the classical action is given by Eq. (\ref{CLAS}), the quadratic
terms by Eq. (\ref{QUAD}) and the cubic and quartic terms by Eq. (\ref{CUB}%
), where

\begin{equation}
S_{c}^{\left( 3+\right) }\left[ \bar{\phi},\delta \phi \right] =\frac{1}{6}%
S_{\alpha \beta \gamma }^{\left( 3\right) }\left[ \bar{\phi}\right] \delta
\phi ^{\alpha }\delta \phi ^{\beta }\delta \phi ^{\gamma }+\frac{1}{24}%
S_{\alpha \beta \gamma \delta }^{\left( 4\right) }\delta \phi ^{\alpha
}\delta \phi ^{\beta }\delta \phi ^{\gamma }\delta \phi ^{\delta }
\end{equation}

The two loops approximation for $\Gamma _2$ reads

\begin{eqnarray}
\Gamma _{2} &=&\frac{1}{24}S_{\alpha \beta \gamma \delta }^{\left( 4\right)
}\left\langle \delta \phi ^{\alpha }\delta \phi ^{\beta }\delta \phi
^{\gamma }\delta \phi ^{\delta }\right\rangle _{2PI}  \nonumber \\
&&+\frac{i}{72}S_{\alpha \beta \gamma }^{\left( 3\right) }\left[ \bar{\phi}%
\right] S_{\alpha ^{\prime }\beta ^{\prime }\gamma ^{\prime }}^{\left(
3\right) }\left[ \bar{\phi}\right] \left\langle \delta \phi ^{\alpha }\delta
\phi ^{\beta }\delta \phi ^{\gamma }\delta \phi ^{\alpha ^{\prime }}\delta
\phi ^{\beta ^{\prime }}\delta \phi ^{\gamma ^{\prime }}\right\rangle _{2PI}
\nonumber \\
&&-\frac{i}{2}f_{\alpha }^{A}T_{1B\beta }^{\alpha }f_{\alpha ^{\prime
}}^{A^{\prime }}T_{1B^{\prime }\beta ^{\prime }}^{\alpha ^{\prime
}}\left\langle \delta \chi _{A}\delta \phi ^{\beta }\delta \omega ^{B}\delta
\chi _{A^{\prime }}\delta \phi ^{\beta ^{\prime }}\delta \omega ^{B^{\prime
}}\right\rangle _{2PI}  \nonumber \\
&=&\frac{1}{4}S_{\alpha \beta \gamma \delta }^{\left( 4\right) }G_{\phi \phi
}^{\alpha \beta }G_{\phi \phi }^{\gamma \delta }+\frac{i}{12}S_{\alpha \beta
\gamma }^{\left( 3\right) }\left[ \bar{\phi}\right] S_{\alpha ^{\prime
}\beta ^{\prime }\gamma ^{\prime }}^{\left( 3\right) }\left[ \bar{\phi}
\right] G_{\phi \phi }^{\alpha \alpha ^{\prime }}G_{\phi \phi }^{\beta \beta
^{\prime }}G_{\phi \phi }^{\gamma \gamma ^{\prime }}  \nonumber \\
&&+\frac{i}{2}f_{\alpha }^{A}T_{1B\beta }^{\alpha }f_{\alpha ^{\prime
}}^{A^{\prime }}T_{1B^{\prime }\beta ^{\prime }}^{\alpha ^{\prime }}G_{\phi
\phi }^{\beta \beta ^{\prime }}G_{\omega \chi A}^{B^{\prime }}G_{\omega \chi
A^{\prime }}^{B}
\end{eqnarray}

We may now write the Schwinger-Dyson equations

\[
0=S_{c,\alpha }+\bar{h}_{A}f_{\alpha }^{A}+\frac{1}{2}S_{c,\alpha \beta
\gamma }\left[ \bar{\phi}\right] G_{\phi \phi }^{\beta \gamma }+\frac{i}{2}%
f_{\beta }^{A}T_{1B\alpha }^{\beta }G_{\chi \omega A}^{B}+\frac{i}{6}
S_{\alpha \beta \gamma \delta }^{\left( 4\right) }S_{\delta ^{\prime }\beta
^{\prime }\gamma ^{\prime }}^{\left( 3\right) }\left[ \bar{\phi}\right]
G_{\phi \phi }^{\delta \delta ^{\prime }}G_{\phi \phi }^{\beta \beta
^{\prime }}G_{\phi \phi }^{\gamma \gamma ^{\prime }} 
\]

\begin{eqnarray*}
0 &=&S_{c,\alpha \beta }-i\left[ \mathbf{G}^{-1}\right] _{\phi \phi \alpha
\beta }+S_{\alpha \beta \gamma \delta }^{\left( 4\right) }G_{\phi \phi
}^{\gamma \delta }+\frac{i}{2}S_{\alpha \delta \gamma }^{\left( 3\right)
}\left[ \bar{\phi}\right] S_{\beta \delta ^{\prime }\gamma ^{\prime
}}^{\left( 3\right) }\left[ \bar{\phi}\right] G_{\phi \phi }^{\delta \delta
^{\prime }}G_{\phi \phi }^{\gamma \gamma ^{\prime }} \\
&&+if_{\gamma }^{A}T_{1B\alpha }^{\gamma }f_{\gamma ^{\prime }}^{A^{\prime
}}T_{1B^{\prime }\beta }^{\gamma ^{\prime }}G_{\omega \chi A}^{B^{\prime
}}G_{\omega \chi A^{\prime }}^{B}
\end{eqnarray*}

\begin{equation}
0=-if_{\alpha }^{A^{\prime }}T_{B}^{\alpha }\left[ \bar{\phi}\right]
+i\left[ \mathbf{G}_{L}^{-1}\right] _{\omega \chi B}^{A^{\prime
}}+if_{\alpha }^{A}T_{1B\beta }^{\alpha }f_{\alpha ^{\prime }}^{A^{\prime
}}T_{1B^{\prime }\beta ^{\prime }}^{\alpha ^{\prime }}G_{\phi \phi }^{\beta
\beta ^{\prime }}G_{\omega \chi A}^{B^{\prime }}  \label{GHOSTPROP}
\end{equation}

So finally we get the equation for the gluon propagator

\begin{eqnarray}
0 &=&S_{c,\alpha \beta }-\frac{1}{\xi }f_{B\alpha }f_{\beta }^{B}-i\left[
G_{\phi \phi }^{-1}\right] _{\alpha \beta }+S_{\alpha \beta \gamma \delta
}^{\left( 4\right) }G_{\phi \phi }^{\gamma \delta }+\frac{i}{2}S_{\alpha
\delta \gamma }^{\left( 3\right) }\left[ \bar{\phi}\right] S_{\beta \delta
^{\prime }\gamma ^{\prime }}^{\left( 3\right) }\left[ \bar{\phi}\right]
G_{\phi \phi }^{\delta \delta ^{\prime }}G_{\phi \phi }^{\gamma \gamma
^{\prime }}  \nonumber \\
&&+if_{\gamma }^{A}T_{1B\alpha }^{\gamma }f_{\gamma ^{\prime }}^{A^{\prime
}}T_{1B^{\prime }\beta }^{\gamma ^{\prime }}G_{\omega \chi A}^{B^{\prime
}}G_{\omega \chi A^{\prime }}^{B}  \label{GLUONPROP}
\end{eqnarray}

The other nontrivial equation is Eq. (\ref{GHOSTPROP}).

Let us assume there is a solution with $\bar{\phi}=0$. Then, since $%
S_{,\alpha }T_{B}^{\alpha }=0$ but also $S_{,\alpha }\left[ 0\right] =0,$ we
have the identities

\[
S_{,\alpha \beta }T_B^\alpha +S_{,\alpha }T_{1B\beta }^\alpha =0\Rightarrow
S_{,\alpha \beta }T_B^\alpha \left[ 0\right] =0 
\]

\[
S_{,\alpha \beta \gamma }T_B^\alpha +S_{,\alpha \beta }T_{1B\gamma }^\alpha
+S_{,\alpha \gamma }T_{1B\beta }^\alpha =0 
\]

\begin{equation}
S_{,\alpha \beta \gamma \delta }T_{B}^{\alpha }+S_{,\alpha \beta \gamma
}T_{1B\delta }^{\alpha }+S_{,\alpha \beta \delta }T_{1B\gamma }^{\alpha
}+S_{,\alpha \gamma \delta }T_{1B\beta }^{\alpha }=0
\end{equation}

\section{The Zinn-Justin equation}

We wish now to derive the Zinn-Justin equation appropiate to $\Gamma _2$.
The key observation is that under a BRST transform within the path integral,
only the source terms are really transformed. Therefore

\begin{equation}
\left\langle \Omega \left[ X^{r}\right] \right\rangle J_{r}+\frac{1}{2}%
\theta ^{rs}\theta ^{s}\left\langle \Omega \left[ X^{r}X^{s}\right]
\right\rangle \mathbf{K}_{rs}=0
\end{equation}

The sources are

\[
J_{r}+\mathbf{K}_{rs}\bar{X}^{s}=-\left[ \frac{\delta }{\delta \bar{X}^{r}}%
\Gamma \right] 
\]

\begin{equation}
\mathbf{K}_{rs}=-2\theta ^{sr}\theta ^{s}\left[ \frac{\delta }{\delta 
\mathbf{G}^{rs}}\Gamma \right]   \label{SOURCE}
\end{equation}
leading to

\begin{eqnarray}
0 &=&\left\langle \Omega \left[ X^{r}\right] \right\rangle \left[ \frac{%
\delta \Gamma }{\delta \bar{X}^{r}}+\mathbf{K}_{rs}\bar{X}^{s}\right] -\frac{%
1}{2}\theta ^{rs}\theta ^{s}\left\langle \Omega \left[ X^{r}X^{s}\right]
\right\rangle \mathbf{K}_{rs}  \nonumber \\
&=&\left\langle \Omega \left[ X^{r}\right] \right\rangle \frac{\delta \Gamma 
}{\delta \bar{X}^{r}}+\theta ^{rs}\theta ^{s}\left[ \left\langle \Omega
\left[ X^{r}\right] \right\rangle \bar{X}^{s}-\frac{1}{2}\left\langle \Omega
\left[ X^{r}X^{s}\right] \right\rangle \right] \mathbf{K}_{rs}  \nonumber \\
&=&\left\langle \Omega \left[ X^{r}\right] \right\rangle \frac{\delta \Gamma 
}{\delta \bar{X}^{r}}+\left[ \left\langle \Omega \left[ X^{r}X^{s}\right]
\right\rangle -2\left\langle \Omega \left[ X^{r}\right] \right\rangle \bar{X}%
^{s}\right] \left[ \frac{\delta }{\delta \mathbf{G}^{rs}}\Gamma \right] 
\label{ZJI}
\end{eqnarray}

For simplicity, we shall assume that all background fields vanish.

Since the Z-J operator has ghost number $1$, it makes no sense to assume
that all quantities with non zero ghost number vanish, as we have done in
the previous section. However, we may still ``turn on'' these quantities one
by one, and thus obtain partial Z-J identities. For example, we get three
identities relating quantities with zero ghost number by requiring that the
coefficients of $\bar{\omega}$ and $G_{\omega \phi }$ vanish (we shall not
investigate the first, as we are assuming no non zero backgrounds, and we
are working throughout with the reduced 2PIEA). This means that we may still
set

\begin{equation}
\bar{\omega}=\bar{\chi}=G_{\omega \omega }=G_{\chi \chi }=G_{\chi \phi
}=G_{\chi h}=0
\end{equation}
and retain only terms linear in $G_{\omega \phi }$ and $G_{\omega h}$. In
this approximation, terms with ghost number neither $0$ or $1$ must vanish
identically, so

\begin{equation}
\left\langle \Omega \left[ \omega ^{D}\right] \right\rangle =\left\langle
\Omega \left[ \omega ^{D}\omega ^{E}\right] \right\rangle =\left\langle
\Omega \left[ h_{A}\omega ^{D}\right] \right\rangle =\left\langle \Omega
\left[ \phi ^{\alpha }\omega ^{D}\right] \right\rangle =\left\langle \Omega
\left[ \chi _{A}\chi _{B}\right] \right\rangle =0
\end{equation}
and

\begin{equation}
\frac{\delta \Gamma }{\delta \omega ^{D}}=\frac{\delta \Gamma }{\delta
G_{\phi \omega }^{\alpha D}}=\frac{\delta \Gamma }{\delta G_{h\omega A}^{D}}=%
\frac{\delta \Gamma }{\delta G_{\omega \omega }^{DE}}=\frac{\delta \Gamma }{%
\delta G_{\chi \chi AB}}=0
\end{equation}
Also, since there are no preferred directions in gauge space, objects with a
single gauge index must vanish out of symmetry, and therefore

\begin{equation}
\left\langle \Omega \left[ \phi ^{\alpha }\right] \right\rangle
=\left\langle \Omega \left[ h^{A}\right] \right\rangle =\left\langle \Omega
\left[ \chi _{A}\right] \right\rangle =\frac{\delta \Gamma }{\delta \bar{\phi%
}^{\alpha }}\left[ 0\right] =0
\end{equation}
Finally, observe that at zero external sources,

\begin{equation}
\frac{\delta \Gamma }{\delta h^{A}}=\frac{\delta \Gamma }{\delta G_{\phi
hB}^{\alpha }}=\frac{\delta \Gamma }{\delta G_{h\chi AB}}=\frac{\delta
\Gamma }{\delta G_{hhAB}}\equiv 0
\end{equation}
In other words, from the terms in Eq. (\ref{ZJI}) we keep the terms in $\phi
\phi ,$ $\phi \chi ,$ and $\chi \omega $ only.

Eq. (\ref{ZJI}) must vanish at the physical point, since each coefficient
vanishes. What is remarkable is that it vanishes identically, even if $%
G_{\phi \omega }^{\alpha A}\neq 0.$ Now $\delta \Gamma /\delta G_{\phi \phi
}^{\alpha \beta }$ and $\delta \Gamma /\delta G_{\omega \chi A}^{B}$ have
ghost number zero, and therefore contain no terms linear in $G_{\phi \omega
}^{\alpha A}.$ We conclude that, to linear order in $G_{\phi \omega
}^{\alpha A},$ we may write

\begin{equation}
\left\langle \Omega \left[ \phi ^{\alpha }\chi _{A}\right] \right\rangle 
\frac{\delta \Gamma }{\delta G_{\phi \chi }^{\alpha A}}\approx 0
\end{equation}

Where $\approx $ means up to terms proportional to the equations of motion.

The transforms involve cubic terms

\begin{equation}
\Omega \left[ \phi ^\alpha \chi _A\right] =T_B^\alpha \left[ \phi \right]
\omega ^B\chi _A+i\phi ^\alpha h_A
\end{equation}

The corresponding expectation values may be computed by adding to the
classical action a new BRST- invariant source term

\begin{equation}
\bar{\kappa}_{\phi \chi \alpha }^{A}\Omega \left[ \phi ^{\alpha }\chi
_{A}\right] 
\end{equation}
so we can write explicitly the action

\begin{equation}
\bar{S}=S_{c}\left[ \phi \right] +h_{A}f_{\alpha }^{A}\phi ^{\alpha }+\frac{%
\xi }{2}h^{A}h_{A}+i\chi _{A}f_{\alpha }^{A}T_{B}^{\alpha }\left[ \phi
\right] \omega ^{B}++\bar{\kappa}_{\phi \chi \alpha }^{A}\left[
T_{B}^{\alpha }\left[ \phi \right] \omega ^{B}\chi _{A}+i\phi ^{\alpha
}h_{A}\right] 
\end{equation}
The quadratic terms are

\begin{equation}
\bar{S}^{\left( 2\right) }=S_{c}^{\left( 2\right) }\left[ \phi \right]
+h_{A}f_{\alpha }^{A}\phi ^{\alpha }+\frac{\xi }{2}h^{A}h_{A}+i\chi
_{A}f_{\alpha }^{A}T_{B}^{\alpha }\left[ 0\right] \omega ^{B}+\bar{\kappa}%
_{\phi \chi \alpha }^{A}\left[ T_{B}^{\alpha }\left[ 0\right] \omega
^{B}\chi _{A}+i\phi ^{\alpha }h_{A}\right] 
\end{equation}
The cubic and quartic terms are

\begin{equation}
\bar{S}^{\left( 3+\right) }=S_{c}^{\left( 3+\right) }\left[ \phi \right]
+i\chi _{A}f_{\alpha }^{A}T_{1B\beta }^{\alpha }\phi ^{\beta }\omega ^{B}+%
\bar{\kappa}_{\phi \chi \alpha }^{A}T_{1B\beta }^{\alpha }\phi ^{\beta
}\omega ^{B}\chi _{A}
\end{equation}
so the missing expectation value is (cfr. Eq. (\ref{TPIEA}))

\begin{equation}
\left\langle \Omega \left[ \phi ^{\alpha }\chi _{A}\right] \right\rangle
=G_{\omega \chi A}^{B}T_{B}^{\alpha }\left[ 0\right] +iG_{\phi hA}^{\alpha }+%
\frac{\delta \Gamma _{2}}{\delta \bar{\kappa}_{\phi \chi \alpha }^{A}}
\end{equation}
and the Z-J identity reads

\begin{equation}
\left[ G_{\omega \chi A}^BT_B^\alpha \left[ 0\right] +iG_{\phi hA}^\alpha +%
\frac{\delta \Gamma _2}{\delta \bar \kappa _{\phi \chi \alpha }^A}\right] 
\frac{\delta \Gamma }{\delta G_{\phi \chi }^{\alpha A}}\approx 0
\end{equation}

Now

\begin{equation}
\frac{\delta \Gamma }{\delta G_{\phi \chi }^{\alpha A}}=\frac{-i}{2}\left[ 
\mathbf{G}_{L}^{-1}\right] _{\phi \chi \alpha A}
\end{equation}
and

\begin{equation}
\left[ \mathbf{G}_{L}^{-1}\right] _{\phi \chi \alpha A}=-\left[ \left[ 
\mathbf{G}_{L}^{-1}\right] _{\phi \phi \alpha \beta }+\frac{i}{\xi }%
f_{\alpha C}f_{\beta }^{C}\right] G_{\omega \phi }^{B\beta }\left[ \mathbf{G}%
_{R}^{-1}\right] _{\chi \omega AB}\approx -\left[ G_{\phi \phi }^{-1}\right]
_{\alpha \beta }G_{\omega \phi }^{C\beta }\left[ \mathbf{G}_{R}^{-1}\right]
_{\chi \omega AC}
\end{equation}
so

\begin{equation}
\left[ T_{C}^{\alpha }\left[ 0\right] \left[ G_{\phi \phi }^{-1}\right]
_{\alpha \beta }-\frac{i}{\xi }\left[ \mathbf{G}_{R}^{-1}\right] _{\chi
\omega AC}f_{\beta }^{A}+\left[ \mathbf{G}_{R}^{-1}\right] _{\chi \omega
C}^{A}\frac{\delta \Gamma _{2}}{\delta \bar{\kappa}_{\phi \chi \alpha }^{A}}%
\left[ G_{\phi \phi }^{-1}\right] _{\alpha \beta }\right] G_{\omega \phi
}^{C\beta }\approx 0
\end{equation}

We must still compute the derivatives w.r.t the external sources. Within the
two loops approximation, we have

\begin{equation}
\frac{\delta \Gamma _{2}}{\delta \bar{\kappa}_{\phi \chi \alpha }^{A}}%
=-\left\langle \left( \chi _{F}f_{\gamma }^{F}T_{1D\delta }^{\gamma }\phi
^{\delta }\omega ^{D}\right) T_{1E\rho }^{\alpha }\phi ^{\rho }\omega
^{E}\chi _{A}\right\rangle _{2PI}=-f_{\gamma }^{F}T_{1D\delta }^{\gamma
}T_{1E\rho }^{\alpha }G_{\chi \omega F}^{E}G_{\chi \omega A}^{D}G_{\phi \phi
}^{\delta \rho }
\end{equation}
so finally

\begin{equation}
\left[ \left[ T_{C}^{\alpha }\left[ 0\right] -f_{\gamma }^{F}T_{1C\delta
}^{\gamma }T_{1E\rho }^{\alpha }G_{\chi \omega F}^{E}G_{\phi \phi }^{\delta
\rho }\right] \left[ G_{\phi \phi }^{-1}\right] _{\alpha \beta }-\frac{i}{%
\xi }\left[ \mathbf{G}_{R}^{-1}\right] _{\chi \omega AC}f_{\beta
}^{A}\right] G_{\omega \phi }^{C\beta }\approx 0  \label{WT1}
\end{equation}

\section{Gauge dependence of the propagators}

To investigate the gauge dependence of the 2PIEA, recall Eqs. (\ref{GFF1}), (%
\ref{GFF2}) and (\ref{GFF3}). Upon a change $\delta F$ in the gauge fermion $%
F$, holding the background fields constant, we get

\begin{equation}
\left. \delta \Gamma \right| _{\bar{X}^{r},\mathbf{G}^{rs}}=\left. \delta
W\right| _{J_{r},\mathbf{K}_{rs}}
\end{equation}

With the same argument as for the 1PIEA, this leads to

\begin{equation}
\left. \delta \Gamma \right| _{\bar{X}^{r},\mathbf{G}^{rs}}=i\left\{
\left\langle \delta F\Omega \left[ X^{r}\right] \right\rangle J_{r}+\frac{1}{%
2}\theta ^{rs}\theta ^{s}\left\langle \delta F\Omega \left[
X^{r}X^{s}\right] \right\rangle \mathbf{K}_{rs}\right\}
\end{equation}

Use Eq. (\ref{SOURCE}) to get

\begin{equation}
\left. \delta \Gamma \right| _{\bar{X}^{r},\mathbf{G}^{rs}}=i\left\{
\left\langle \delta F\Omega \left[ X^{r}\right] \right\rangle \frac{\delta
\Gamma }{\delta \bar{X}^{r}}+\left[ \left\langle \delta F\Omega \left[
X^{r}X^{s}\right] \right\rangle -2\left\langle \delta F\Omega \left[
X^{r}\right] \right\rangle \bar{X}^{s}\right] \left[ \frac{\delta }{\delta 
\mathbf{G}^{rs}}\Gamma \right] \right\}
\end{equation}

As before, we shall assume that all background fields vanish and that at
such a point $\Gamma _{,r}$ vanishes identically, so the above expression
simplifies to

\begin{equation}
\left. \delta \Gamma \right| _{\bar{X}^{r},\mathbf{G}^{rs}}=Y^{rs}\frac{%
\delta }{\delta \mathbf{G}^{rs}}\Gamma ;\qquad Y^{rs}=i\left\langle \delta
F\Omega \left[ X^{r}X^{s}\right] \right\rangle
\end{equation}

At the physical point, the Schwinger - Dyson equations now read

\begin{equation}
\frac{\delta }{\delta \mathbf{G}^{tu}}\Gamma +Y^{rs}\frac{\delta ^{2}}{%
\delta \mathbf{G}^{tu}\delta \mathbf{G}^{rs}}\Gamma =0
\end{equation}

Of course, the solution is now $\mathbf{G}^{tu}+\delta \mathbf{G}^{tu}$, so

\begin{equation}
\left( \frac{\delta ^{2}}{\delta \mathbf{G}^{tu}\delta \mathbf{G}^{rs}}%
\Gamma \right) \left[ \delta \mathbf{G}^{rs}+Y^{rs}\right] =0
\end{equation}

Since the Hessian is suppossed to be invertible, we must have $\delta 
\mathbf{G}^{rs}=-Y^{rs}$.

Let us also assume that all propagators with non zero ghost number vanish.
Thus we are only concerned with

\[
\left. \delta \Gamma \right| _{\bar{X}^{r},\mathbf{G}^{rs}}=\left\langle
\delta F\Omega \left[ \phi ^{\alpha }\phi ^{\beta }\right] \right\rangle 
\frac{\delta \Gamma }{\delta G_{\phi \phi }^{\alpha \beta }}+2\left\langle
\delta F\Omega \left[ \chi _{A}\omega ^{D}\right] \right\rangle \frac{\delta
\Gamma }{\delta G_{\omega \chi A}^{D}} 
\]

\[
\Omega \left[ \phi ^{\alpha }\phi ^{\beta }\right] =T_{A}^{\alpha }\left[
\phi \right] \omega ^{A}\phi ^{\beta }+\phi ^{\alpha }T_{A}^{\beta }\left[
\phi \right] \omega ^{A} 
\]

\[
\Omega \left[ \chi _{A}\omega ^{D}\right] =ih_{A}\omega ^{D}-\frac{1}{2}\chi
_{A}C_{\;CB}^{D}\omega ^{C}\omega ^{B} 
\]

\begin{equation}
\delta F=-i\chi _{A}\left\{ \delta f^{A}\left[ \phi \right] +\frac{1}{2}%
\delta \xi h^{A}\right\}
\end{equation}

As before, we compute the corresponding expectation values by adding
suitable sources to the action. These sources correspond to four and five -
legged vertices. Now, recall the conventional argument that

\[
l-1=i-\sum v_{n} 
\]

\begin{equation}
2i-\sum nv_{n}=0
\end{equation}

In our case, we have vertices with three, four or five legs, and $l=2,$ so

\[
2i-3v_{3}-4v_{4}-5v_{5}=0 
\]

\[
i=1+v_{3}+v_{4}+v_{5} 
\]

\begin{equation}
2-v_{3}-2v_{4}-3v_{5}=0
\end{equation}

We are interested in the case where $v_{5}=0$ or $1$. In the latter, we get
an impossibility, so we must have $v_{5}=0.$ Since we may discard the case
when $v_{4}=0$ too, we must have $v_{4}=1$, $v_{3}=0,$ $i=2$. In other
words, to two loops, the only contributions to the expectation values are
those where there are only four fields involved, and these contract among
themselves.

In conclusion,

\begin{eqnarray}
\delta G_{\phi \phi }^{\alpha \beta } &=&-T_A^\alpha \left[ 0\right] \left[
\delta f_\gamma ^B\left\langle \chi _B\phi ^\gamma \omega ^A\phi ^\beta
\right\rangle +\frac 12\delta \xi \left\langle \chi _Bh^B\omega ^A\phi
^\beta \right\rangle \right] +\left( \alpha \leftrightarrow \beta \right) 
\nonumber \\
&=&-T_A^\alpha \left[ 0\right] \left[ \delta f_\gamma ^B-\frac{\delta \xi }{%
2\xi }f_\gamma ^B\right] \left\langle \chi _B\phi ^\gamma \omega ^A\phi
^\beta \right\rangle +\left( \alpha \leftrightarrow \beta \right)  \nonumber
\\
&=&-T_A^\alpha \left[ 0\right] \left[ \delta f_\gamma ^B-\frac{\delta \xi }{%
2\xi }f_\gamma ^B\right] G_{\phi \phi }^{\beta \gamma }G_{\chi \omega
B}^A+\left( \alpha \leftrightarrow \beta \right)
\end{eqnarray}

We may also write

\begin{equation}
\delta G_{\phi \phi \alpha \beta }^{-1}=G_{\phi \phi \alpha \delta
}^{-1}T_{A}^{\delta }\left[ 0\right] \left[ \delta f_{\beta }^{B}-\frac{%
\delta \xi }{2\xi }f_{\beta }^{B}\right] G_{\chi \omega B}^{A}+\left( \alpha
\leftrightarrow \beta \right)
\end{equation}

This shows that the zeroes of the inverse propagator are gauge invariant. To
lowest order, the Z-J equation

\begin{equation}
\left[ \left[ T_{C}^{\alpha }\left[ 0\right] -f_{\gamma }^{F}T_{1C\delta
}^{\gamma }T_{1E\rho }^{\alpha }G_{\chi \omega F}^{E}G_{\phi \phi }^{\delta
\rho }\right] \left[ G_{\phi \phi }^{-1}\right] _{\alpha \beta }-\frac{i}{%
\xi }\left[ \mathbf{G}_{R}^{-1}\right] _{\chi \omega AC}f_{\beta
}^{A}\right] G_{\omega \phi }^{C\beta }\approx 0
\end{equation}
implies

\begin{eqnarray}
\delta G_{\phi \phi \alpha \beta }^{-1} &=&\frac{i}{\xi }\left[ \mathbf{G}%
_{R}^{-1}\right] _{\chi \omega AC}f_{\alpha }^{C}\left[ \delta f_{\beta
}^{B}-\frac{\delta \xi }{2\xi }f_{\beta }^{B}\right] G_{\chi \omega
B}^{A}+\left( \alpha \leftrightarrow \beta \right)  \nonumber \\
&=&\frac{i}{\xi }\left[ f_{B\alpha }\delta f_{\beta }^{B}+f_{B\beta }\delta
f_{\alpha }^{B}-\frac{\delta \xi }{\xi }f_{\beta }^{B}f_{B\alpha }\right]
\label{GAUGEDEPEN}
\end{eqnarray}

This is the result we wanted to show.

\section{Propagator structure}

We shall try and apply the foregoing to clarify the structure of the 2-loop
propagators. Let us begin with the equation for the ghost propagator(cfr.
Eq. (\ref{GHOSTPROP}))

\begin{equation}
\left[ \mathbf{G}_{L}^{-1}\right] _{\omega \chi B}^{A^{\prime }}=f_{\alpha
}^{A^{\prime }}\left[ T_{B}^{\alpha }\left[ \bar{\phi}\right] -f_{\alpha
^{\prime }}^{A}T_{1B\beta }^{\alpha ^{\prime }}T_{1B^{\prime }\beta ^{\prime
}}^{\alpha }G_{\phi \phi }^{\beta \beta ^{\prime }}G_{\omega \chi
A}^{B^{\prime }}\right]
\end{equation}

This suggests defining

\begin{equation}
P_{L\beta }^{\alpha }=\left[ T_{B}^{\alpha }\left[ \bar{\phi}\right]
-f_{\alpha ^{\prime }}^{A}T_{1B\gamma }^{\alpha ^{\prime }}T_{1B^{\prime
}\beta ^{\prime }}^{\alpha }G_{\phi \phi }^{\gamma \beta ^{\prime
}}G_{\omega \chi A}^{B^{\prime }}\right] G_{\omega \chi C}^{B}f_{\beta }^{C}
\end{equation}
which is a projection operator

\begin{equation}
P_{L\beta }^{\alpha }P_{L\gamma }^{\beta }=P_{L\gamma }^{\alpha }
\end{equation}

Now consider the Takahashi-Ward identity (cfr. Eq. (\ref{WT1})) 
\begin{equation}
\left[ T_{B}^{\alpha }\left[ 0\right] -T_{1B^{\prime }\beta ^{\prime
}}^{\alpha }f_{\alpha ^{\prime }}^{A}T_{1B\gamma }^{\alpha ^{\prime
}}G_{\chi \omega A}^{B^{\prime }}G_{\phi \phi }^{\gamma \beta ^{\prime
}}\right] \left[ G_{\phi \phi }^{-1}\right] _{\alpha \lambda }-\frac{i}{\xi }%
\left[ \mathbf{G}_{R}^{-1}\right] _{\chi \omega AB}f_{\lambda }^{A}\approx 0
\end{equation}
It becomes

\begin{equation}
P_{L\beta }^{\alpha }\left[ G_{\phi \phi }^{-1}\right] _{\alpha \lambda }-%
\frac{i}{\xi }\left[ \mathbf{G}_{R}^{-1}\right] _{\chi \omega AB}f_{\lambda
}^{A}G_{\omega \chi C}^{B}f_{\beta }^{C}\approx 0
\end{equation}
That is

\begin{equation}
P_{L\beta }^{\alpha }\left[ G_{\phi \phi }^{-1}\right] _{\alpha \lambda }-%
\frac{i}{\xi }f_{\lambda }^{A}f_{A\beta }\approx 0
\end{equation}
Therefore

\begin{equation}
\left[ G_{\phi \phi }^{-1}\right] _{\alpha \lambda }=\left[ G_{T\phi \phi
}^{-1}\right] _{\alpha \lambda }+\frac{i}{\xi }f_{\lambda }^{A}f_{A\alpha
},\qquad P_{L\alpha }^{\gamma }\left[ G_{T\phi \phi }^{-1}\right] _{\gamma
\lambda }\approx 0
\end{equation}
Comparing with the gauge-dependence identity Eq. (\ref{GAUGEDEPEN}) we see
that the transverse part $\left[ G_{T\phi \phi }^{-1}\right] _{\gamma
\lambda }$ is gauge-fixing independent. Comparing with the equation of
motion Eq. (\ref{GLUONPROP}) we get

\begin{eqnarray}
0 &=&S_{c,\alpha \beta }-i\left[ G_{T\phi \phi }^{-1}\right] _{\gamma \beta
}+S_{\alpha \beta \gamma \delta }^{\left( 4\right) }G_{\phi \phi }^{\gamma
\delta }+\frac{i}{2}S_{\alpha \delta \gamma }^{\left( 3\right) }\left[ \bar{%
\phi}\right] S_{\beta \delta ^{\prime }\gamma ^{\prime }}^{\left( 3\right)
}\left[ \bar{\phi}\right] G_{\phi \phi }^{\delta \delta ^{\prime }}G_{\phi
\phi }^{\gamma \gamma ^{\prime }}  \nonumber \\
&&+if_{\gamma }^{A}T_{1B\alpha }^{\gamma }f_{\gamma ^{\prime }}^{A^{\prime
}}T_{1B^{\prime }\beta }^{\gamma ^{\prime }}G_{\omega \chi A}^{B^{\prime
}}G_{\omega \chi A^{\prime }}^{B}
\end{eqnarray}

The decomposition of the inverse propagators leads to a related
decomposition of the propagators themselves. We have

\begin{equation}
\frac{i}{\xi }f_{\gamma }^{A}f_{A\lambda }G_{\phi \phi }^{\lambda \beta
}=P_{L\gamma }^{\beta }
\end{equation}
Therefore

\begin{equation}
G_{\phi \phi }^{\lambda \beta }=G_{T\phi \phi }^{\lambda \beta }-i\xi
L_{C}^{\lambda }L^{C\beta },
\end{equation}
where

\begin{equation}
L_{C}^{\lambda }=\left[ T_{B}^{\lambda }\left[ \bar{\phi}\right] -f_{\alpha
^{\prime }}^{A}T_{1B\gamma }^{\alpha ^{\prime }}T_{1B^{\prime }\beta
^{\prime }}^{\lambda }G_{\phi \phi }^{\gamma \beta ^{\prime }}G_{\omega \chi
A}^{B^{\prime }}\right] G_{\omega \chi C}^{B},
\end{equation}
and

\begin{equation}
f_{A\lambda }G_{T\phi \phi }^{\lambda \beta }=0
\end{equation}

Observe that

\begin{equation}
P_{L\beta }^{\alpha }=L_{C}^{\alpha }f_{\beta }^{C}
\end{equation}
and so

\begin{equation}
P_{L\beta }^{\alpha }P_{L\gamma }^{\beta }=L_{C}^{\alpha }f_{\beta
}^{C}L_{D}^{\beta }f_{\gamma }^{D}\Rightarrow f_{\beta }^{C}L_{D}^{\beta
}=\delta _{D}^{C}
\end{equation}
We now have

\begin{equation}
\left[ \left[ G_{T\phi \phi }^{-1}\right] _{\alpha \lambda }+\frac{i}{\xi }%
f_{\lambda }^{A}f_{A\alpha }\right] \left[ G_{T\phi \phi }^{\lambda \beta
}-i\xi L_{C}^{\lambda }L^{C\beta }\right] =\delta _{\alpha }^{\beta }
\end{equation}
so

\begin{equation}
\left[ G_{T\phi \phi }^{-1}\right] _{\alpha \lambda }G_{T\phi \phi
}^{\lambda \beta }=\delta _{\alpha }^{\beta }-P_{L\alpha }^{\beta }
\end{equation}

\section{Appendix: Grassmann calculus}

To be definite, we shall collect several known results concerning Grassmann
calculus, which are necessary for the evaluation of the 2PIEA. For more
details, we refer the reader to the monographs by Berezin \cite{Berezin},
DeWitt \cite{DeWitt2} and Negele and Orland \cite{NO98} .

Let us consider a set of independent Grassmann variables $\xi ^{u}.$ We
define the left derivative $\partial /\partial \xi ^{u}$ from the properties 
$\left\{ \partial /\partial \xi ^{u},\xi ^{v}\right\} =\delta _{u}^{v},$ $%
\partial 1/\partial \xi ^{u}=0.$

We also define Grassmann integrals

\begin{equation}
\int d\xi ^{u}\;\xi ^{v}=\delta ^{uv},\qquad \int d\xi ^{u}\;1=0
\end{equation}

Observe that if $\xi =a\eta $, then $d\xi =a^{-1}d\eta $ (this follows from $%
\int d\xi $ $\xi =\int d\eta $ $\eta $). This allows us to prove the basic
Gaussian integration formula

\begin{equation}
\int d^{n}\xi \;exp\left\{ \frac{-1}{2}\xi ^{u}M_{uv}\xi ^{v}\right\}
\propto \sqrt{det\,M}
\end{equation}

Let us now consider the more general expression

\begin{equation}
\int d^{n}\xi \;exp\left\{ \frac{-1}{2}\xi ^{u}M_{uv}\xi ^{v}+i\theta
_{v}\xi ^{v}\right\} 
\end{equation}
where the $\theta $'s are themselves Grassmann. Then, since $M$ must be
antisymmetric,

\begin{equation}
\xi ^{u}M_{uv}\xi ^{v}-2i\theta _{v}\xi ^{v}=\left( \xi ^{u}-i\left(
M^{-1}\right) ^{uw}\theta _{w}\right) M_{uv}\left( \xi ^{v}-i\left(
M^{-1}\right) ^{vx}\theta _{x}\right) -\theta _{u}\left( M^{-1}\right)
^{uv}\theta _{v}
\end{equation}
so

\begin{equation}
\int d^{n}\xi \;exp\left\{ \frac{-1}{2}\xi ^{u}M_{uv}\xi ^{v}+i\theta
_{v}\xi ^{v}\right\} \propto \sqrt{det\,M}\;exp\left\{ \frac{1}{2}\theta
_{u}\left( M^{-1}\right) ^{uv}\theta _{v}\right\}
\end{equation}

By differentiation, we get

\begin{eqnarray}
\int d^{n}\xi \;\xi ^{u}\xi ^{v}exp\left\{ \frac{-1}{2}\xi ^{u}M_{uv}\xi
^{v}\right\} &=&\frac{\partial ^{2}}{\partial \theta ^{v}\partial \theta ^{u}%
}\int d^{n}\xi \;exp\left\{ \frac{-1}{2}\xi ^{u}M_{uv}\xi ^{v}+i\theta
_{v}\xi ^{v}\right\} _{\theta =0}  \nonumber \\
&\propto &\sqrt{det\,M}\;\left( M^{-1}\right) ^{uv}
\end{eqnarray}

This also follows from an integration by parts

\begin{eqnarray}
\int d^{n}\xi \;\xi ^{u}\xi ^{v}exp\left\{ \frac{-1}{2}\xi ^{u}M_{uv}\xi
^{v}\right\} &=&\left( M^{-1}\right) ^{vw}\int d^{n}\xi \;\xi ^{u}M_{wx}\xi
^{x}exp\left\{ \frac{-1}{2}\xi ^{u}M_{uv}\xi ^{v}\right\}  \nonumber \\
&=&\left( M^{-1}\right) ^{wv}\int d^{n}\xi \;\xi ^{u}\frac{\partial }{%
\partial \xi ^{w}}exp\left\{ \frac{-1}{2}\xi ^{u}M_{uv}\xi ^{v}\right\} 
\nonumber \\
&=&\left( M^{-1}\right) ^{wv}\int d^{n}\xi \;\left[ \frac{\partial \xi ^{u}}{%
\partial \xi ^{w}}\right] exp\left\{ \frac{-1}{2}\xi ^{u}M_{uv}\xi
^{v}\right\}
\end{eqnarray}

Observe that integrating by parts a Grassmann variable does not change the
sign.

We are interested in quadratic forms involving both Grassmann and normal
variables. Let $X^{r}=\left( x^{\alpha },\xi ^{u}\right) $, where the $x$'s
are normal, and consider the expression :$X^{r}\mathbf{M}_{rs}X^{s}:=x^{%
\alpha }H_{\alpha \beta }x^{\beta }+\xi ^{u}M_{uv}\xi ^{v}+2\xi
^{v}N_{v\alpha }x^{\alpha }$. We shall call the object

\begin{equation}
\left( 
\begin{array}{ll}
H_{\alpha \beta } & -N_{v\alpha } \\ 
N_{u\beta } & M_{uv}
\end{array}
\right) 
\end{equation}
a supermatrix. Observe that $H$ is normal and symmetric, $M$ is normal
antisymmetric and $N$ is Grassmann. Suppose the supermatrix $\mathbf{M}$ has
a (right) inverse $\mathbf{M}_{R}^{-1}$ with elements

\begin{equation}
\left( 
\begin{array}{ll}
\bar{H}^{\alpha \beta } & \bar{N}^{v\alpha } \\ 
-\bar{N}^{u\beta } & \bar{M}^{uv}
\end{array}
\right) 
\end{equation}
Then

\begin{eqnarray}
H_{\alpha \beta }\bar{H}^{\beta \gamma }+N_{v\alpha }\bar{N}^{v\gamma }
&=&\delta _{\alpha }^{\gamma }  \nonumber \\
H_{\alpha \beta }\bar{N}^{v\beta }-N_{u\alpha }\bar{M}^{uv} &=&0  \nonumber
\\
N_{u\beta }\bar{H}^{\beta \gamma }-M_{uv}\bar{N}^{v\gamma } &=&0  \nonumber
\\
N_{u\beta }\bar{N}^{v\beta }+M_{uw}\bar{M}^{wv} &=&\delta _{u}^{v}
\end{eqnarray}
Therefore

\begin{equation}
\bar{N}^{v\beta }=\left( H^{-1}\right) ^{\beta \alpha }N_{u\alpha }\bar{M}%
^{uv}=\left( M^{-1}\right) ^{vu}N_{u\gamma }\bar{H}^{\gamma \beta }
\end{equation}

\begin{equation}
\bar{H}^{\alpha \beta }=\left[ H_{\alpha \beta }+N_{\alpha v}\left(
M^{-1}\right) ^{vu}N_{\beta u}\right] ^{-1}
\end{equation}

\begin{equation}
\bar{M}^{uv}=\left[ M_{uv}+N_{\beta u}\left( H^{-1}\right) ^{\beta \alpha
}N_{\alpha v}\right] ^{-1}
\end{equation}

We wish to check that both representations of $\bar{N}^{v\beta }$ are
equivalent. This means that

\begin{equation}
\left( H^{-1}\right) ^{\beta \alpha }N_{u\alpha }\bar{M}^{uv}=\left(
M^{-1}\right) ^{vu}N_{u\gamma }\bar{H}^{\gamma \beta }
\end{equation}
Multiply both sides by $\left[ H_{\beta \delta }+N_{\beta v}\left(
M^{-1}\right) ^{vu}N_{\delta u}\right] $

\begin{equation}
\left( H^{-1}\right) ^{\beta \alpha }N_{\alpha u}\bar{M}^{uv}\left[ H_{\beta
\delta }+N_{\beta x}\left( M^{-1}\right) ^{xy}N_{\delta y}\right] =\left(
M^{-1}\right) ^{vw}N_{\delta w}
\end{equation}
Multiply both sides by $\left[ M_{vz}+N_{\beta v}\left( H^{-1}\right)
^{\beta \alpha }N_{\alpha z}\right] $

\begin{equation}
-\left( H^{-1}\right) ^{\beta \alpha }N_{\alpha z}\left[ H_{\beta \delta
}+N_{\beta x}\left( M^{-1}\right) ^{xy}N_{\delta y}\right] =\left(
M^{-1}\right) ^{vw}N_{\delta w}\left[ M_{vz}+N_{\beta v}\left( H^{-1}\right)
^{\beta \alpha }N_{\alpha z}\right] 
\end{equation}
We must check that

\begin{equation}
\left( H^{-1}\right) ^{\beta \alpha }N_{\alpha z}N_{\beta x}\left(
M^{-1}\right) ^{xy}N_{\delta y}=\left( M^{-1}\right) ^{wv}N_{\delta
w}N_{\beta v}\left( H^{-1}\right) ^{\beta \alpha }N_{\alpha z}
\end{equation}
Which is true, because the $N$'s are Grassmann and $M$ antisymmetric.

By the way, the left inverse is given by

\begin{equation}
\mathbf{M}_L^{-1}=\left( 
\begin{array}{ll}
\bar H^{\alpha \beta } & -\bar N^{v\alpha } \\ 
\bar N^{u\beta } & \bar M^{uv}
\end{array}
\right)
\end{equation}

Let us suppose we use a supermatrix to implement a change of variables

\begin{equation}
X^{r}=\mathbf{M}_{\;s}^{r}Y^{s}
\end{equation}
Spelled out in full,

\begin{eqnarray}
x^{\alpha } &=&H^{\alpha \beta }y_{\beta }-N^{v\alpha }\eta _{v}  \nonumber
\\
\xi ^{u} &=&N^{u\beta }y_{\beta }+M^{uv}\eta _{v}
\end{eqnarray}
From the first equation

\begin{equation}
y=H^{-1}\left( x+N\eta \right) 
\end{equation}
So

\begin{equation}
\xi -NH^{-1}x=\left( M+NH^{-1}N\right) \eta 
\end{equation}
It follows that

\begin{eqnarray}
d^{n}xd^{m}\xi  &=&d^{n}xd^{m}\left( \xi -NH^{-1}x\right) =\mathrm{det}%
\left( M+NH^{-1}N\right) ^{-1}d^{n}xd^{m}\eta   \nonumber \\
&=&\mathrm{det}\left( M+NH^{-1}N\right) ^{-1}\mathrm{det}\left( H\right)
\;d^{n}yd^{m}\eta 
\end{eqnarray}
So, if we define

\begin{equation}
\mathrm{sdet}\mathbf{M}=\mathrm{det}\left[ \mathbf{M}^{\alpha \beta }\right] 
\mathrm{det}\left[ \left( \mathbf{M}_{R}^{-1}\right) ^{uv}\right] 
\end{equation}
then $X=\mathbf{M}Y$ implies $dX=\left( \mathrm{sdet}\mathbf{M}\right) dY.$
Observe that

\begin{equation}
\mathrm{sdet}\mathbf{M}_{R}^{-1}=\left( \mathrm{sdet}\mathbf{M}\right) ^{-1}
\end{equation}
Indeed

\begin{equation}
\mathrm{sdet}\mathbf{M}_{R}^{-1}=\left[ \mathrm{det}\left( H+NM^{-1}N\right)
\right] ^{-1}\mathrm{det}\left( M\right) 
\end{equation}
so we need to show that

\begin{equation}
\mathrm{det}\left( M+NH^{-1}N\right) ^{-1}\mathrm{det}\left( H\right)
=\left[ \mathrm{det}\left( H+NM^{-1}N\right) \right] \left[ \mathrm{det}%
\left( M\right) \right] ^{-1}
\end{equation}
or

\begin{equation}
\mathrm{det}\left( 1+M^{-1}NH^{-1}N\right) ^{-1}=\mathrm{det}\left(
1+H^{-1}NM^{-1}N\right)
\end{equation}

This follows from the $N$'s being Grassmann. To show this, observe that we
can allways diagonalize $H$ and reduce $M$ to $2\times 2$ blocks, so we may
assume that $H^{-1}=h$ is an scalar, $N=\left( \theta _{1},\theta
_{2}\right) $ and

\begin{equation}
M^{-1}=\left( 
\begin{array}{ll}
0 & m \\ 
-m & 0
\end{array}
\right) 
\end{equation}
Observe that

\begin{equation}
NH^{-1}N=h\left( 
\begin{array}{ll}
0 & \theta _{1}\theta _{2} \\ 
-\theta _{1}\theta _{2} & 0
\end{array}
\right) 
\end{equation}
while $NM^{-1}N=2m\theta _{1}\theta _{2}$, so everything reduces to

\begin{equation}
\left( 1-mh\theta _{1}\theta _{2}\right) ^{-2}=1+2mh\theta _{1}\theta _{2}
\end{equation}
which is true because only the linear term in the Taylor development of the
left hand side survives.

We can now prove the Gaussian formula

\begin{equation}
\int dX\;\mathrm{exp}\left\{ \frac{-1}{2}X\mathbf{M}X\right\} \propto \left( 
\mathrm{sdet}\mathbf{M}\right) ^{-1/2}
\end{equation}
Let us now compute

\begin{equation}
\left\langle X^{r}X^{s}\right\rangle \equiv \frac{\int dX\;X^{r}X^{s}\mathrm{%
exp}\left\{ \frac{-1}{2}X\mathbf{M}X\right\} }{\int dX\;\mathrm{exp}\left\{ 
\frac{-1}{2}X\mathbf{M}X\right\} }
\end{equation}
Observe that

\begin{equation}
X^{r}X^{s}=X^{r}\left( \mathbf{M}_{L}^{-1}\right) ^{st}\mathbf{M}%
_{tu}X^{u}=\theta ^{rs}\theta ^{rt}\left( \mathbf{M}_{L}^{-1}\right)
^{st}X^{r}\mathbf{M}_{tu}X^{u}
\end{equation}
On the other hand

\begin{eqnarray}
\frac{\partial }{\partial X^{t}}\mathrm{exp}\left\{ \frac{-1}{2}X\mathbf{M}%
X\right\}  &=&\frac{-1}{2}\left[ \mathbf{M}_{tu}X^{u}+\theta ^{t}X^{u}%
\mathbf{M}_{ut}\right] \mathrm{exp}\left\{ \frac{-1}{2}X\mathbf{M}X\right\} 
\nonumber \\
&=&-\mathbf{M}_{tu}X^{u}\mathrm{exp}\left\{ \frac{-1}{2}X\mathbf{M}X\right\} 
\end{eqnarray}
Finally

\begin{equation}
\frac{\partial }{\partial X^{t}}X^{r}F=\delta _{t}^{r}F+\theta ^{rt}X^{r}%
\frac{\partial F}{\partial X^{t}}
\end{equation}
Puting all together

\begin{equation}
\left\langle X^{r}X^{s}\right\rangle =\theta ^{rs}\left( \mathbf{M}%
_{L}^{-1}\right) ^{sr}=\left( \mathbf{M}_{R}^{-1}\right) ^{rs}
\end{equation}
This shows in particular that

\begin{equation}
\frac \partial {\partial \mathbf{M}_{rs}}\ln \left( \mathrm{sdet}\mathbf{M}%
\right) =\theta ^r\theta ^{rs}\left( \mathbf{M}_R^{-1}\right) ^{rs}
\end{equation}

\textbf{Acknowledgements} It is a pleasure to acknowledge many discussions
with B-L. Hu. This work has been supported in part by CONICET, UBA,
Fundaci\'{o}n Antorchas and ANPCyT.

\end{document}